\UseRawInputEncoding 
\documentclass[journal]{IEEEtran}
\usepackage{graphicx}
\usepackage{amsmath,amssymb}
\usepackage[colorlinks=true, pdfstartview=FitV, linkcolor=blue, citecolor=blue, urlcolor=blue]{hyperref}
\usepackage{float}
\restylefloat{table}
\usepackage[english, footnote, final, nomargin, marginclue]{fixme}
\allowdisplaybreaks

\usepackage{ulem} 
\usepackage{xcolor}

\begin{document}
\bstctlcite{BSTcontrol}
\title{To Trust or Not to Trust: Evolutionary Dynamics of an Asymmetric N-player Trust Game}

\author{Ik~Soo~Lim and Naoki~Masuda
%13 Feburary, 2021 % <-this % stops a space
\thanks{Ik~Soo~Lim  is with 
School of Computing and Mathematical Sciences, University of Greenwich, London SE10 9LS, U.K. 
(e-mail: i.lim@gre.ac.uk)
\\
Naoki~Masuda is with 
Department of Mathematics and Computational and Data-Enabled Science and Engineering Program, 
State University of New York at Buffalo, USA
}% <-this % stops a space
\thanks{Manuscript received XXXX; revised YYYY.}}

\markboth{
}%
{Shell \MakeLowercase{\textit{et al.}}: Bare Demo of IEEEtran.cls for IEEE Journals}

\maketitle


% \nm{I commented out {\textbackslash}pdfoutput=1 on line 2 to enable compilation in my environment.}

\begin{abstract}
		
Trusting others and reciprocating the received trust with trustworthy actions are fundaments of economic and social interactions. The trust game (TG) is widely used for studying trust and trustworthiness and entails a sequential interaction between two players, an investor and a trustee. 
It requires at least two strategies or options for an investor (e.g.\,to trust versus not to trust a trustee). According to the evolutionary game theory, the antisocial strategies (e.g.\,not to trust) evolve such that the investor and trustee end up with lower payoffs than those that they would get with the prosocial strategies (e.g.\,to trust).
A generalisation of the TG to a multiplayer (i.e.\,more than two players) TG was recently proposed. However, its outcomes hinge upon two assumptions that various real situations may substantially deviate from: (i) investors are forced to trust trustees and (ii)
investors can turn into trustees by imitation
and vice versa. We propose an asymmetric multiplayer TG that allows investors not to trust and prohibits  
the imitation between players of different roles; instead, investors learn from other investors and the same for trustees.
We show that the evolutionary game dynamics of the proposed TG qualitatively depends on the nonlinearity of the payoff function and the amount of incentives collected from and distributed to players through an institution. 
We also show that incentives given to trustees can be useful and  sufficient to cost-effectively promote trust and trustworthiness among self-interested players.
\end{abstract}


% \fi
\IEEEpeerreviewmaketitle

\section{Introduction}

The evolution of pro-social behaviours among self-interested individuals has been 
a focus of research across disciplines.
For instance, the evolution of cooperation in social dilemma situations such as the Prisoner's Dilemma (PD) 
and its  $N$-player generalisation, the Public Goods Game (PGG),
has attracted lots of attention
\cite{Li:2017kx}\cite{Chiong:2012uq}\cite{nowak2006five}\cite{hauert2006synergy}\cite{Van-Segbroeck:2009uq}\cite{Sasaki:2012rz}. 
Evolutionary game theory provides a theoretical framework with which to study the evolution of strategies or behaviours  among self-interested individuals in these social dilemmas or other situations, in which successful strategies or genes are spread  by fitness-dependent reproduction and imitation \cite{SMITH:1973aa}\cite{Taylor:1978aa}.
It has also been widely used for applications such as modelling the propagation of competing technologies and policies for green supply chain management
\cite{Bolluyt:2019um}\cite{Long:2021wt}.
	
\begin{figure}[t!]  
\begin{center}  
\includegraphics[width=0.48\textwidth]{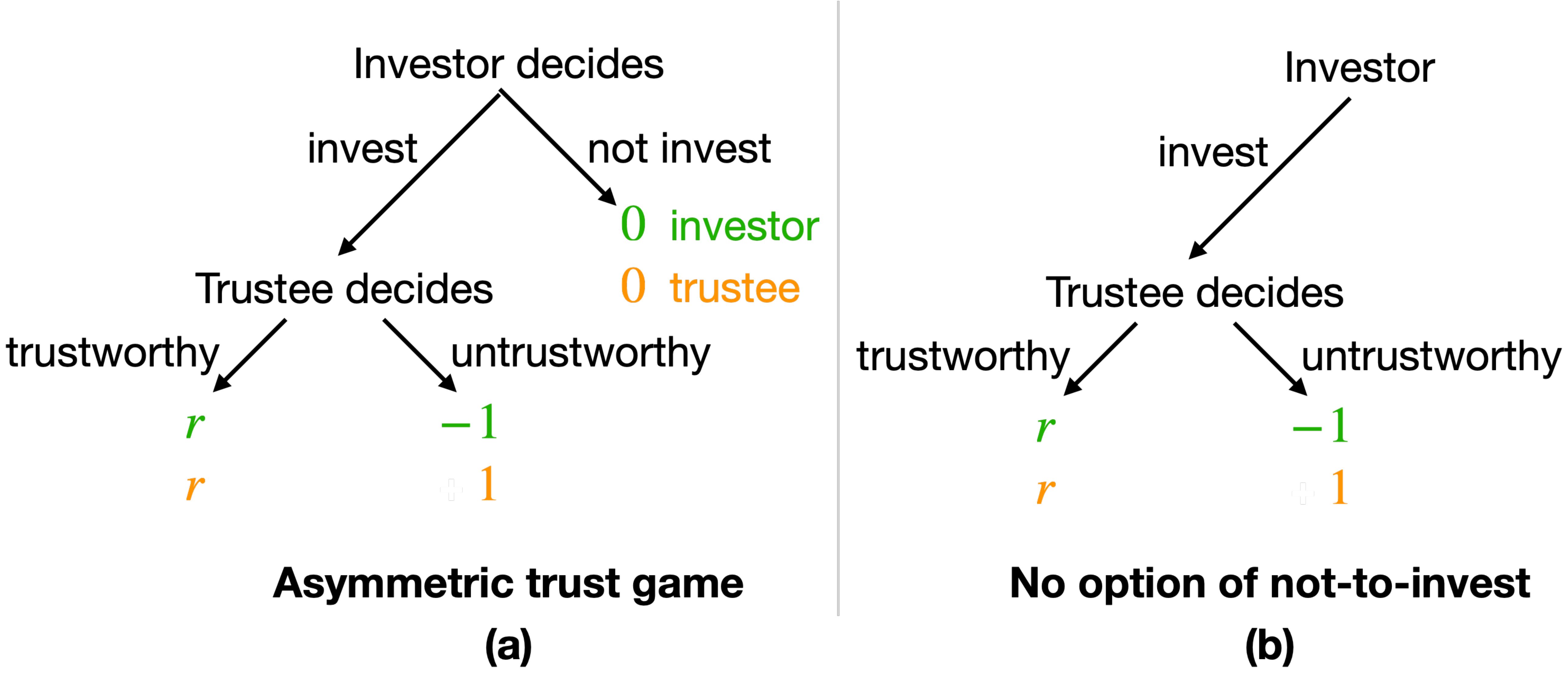}             
\end{center}  
\caption{
Two-player binary TGs.
(a) Game tree of the asymmetric two-player binary TG,
referred to as a general two-player TG1G in Ref.~\cite{Abbass:2016aa}, in which the role of each player is fixed. 
The payoffs of an investor are shown in green. Those of a trustee are shown in orange.
Adapted from Ref.~\cite{Masuda:2012aa}.
We generalise this game to an $N$-player game in this article.
(b) Game tree of the two-player binary TG that is used for the generalisation to the NTG in Ref.~\cite{Abbass:2016aa}.
This game does not allow an investor not to invest. 
In both (a) and (b), we require $0<r<1$, where $r$ represents the relative productivity of the prosocial strategies. 
}    
\label{fig_trust_game}  
\end{figure}

Non-simultaneous or sequential interactions between two players are common in many situations
such as buyer-seller interactions, whereas the PD and PGG are concerned with simultaneous interactions.
Non-simultaneous interactions yield a problem of trust
in the sense that  the decision by one of two players (e.g.\,a buyer) can make oneself vulnerable to potential exploitation by the other (e.g.\,a seller) \cite{Bravo:2008aa}. 
In such situations, higher levels of trusting in others and reciprocating the received trust with trustworthy actions  have been associated with more efficient judicial systems, higher quality in government bureaucracies, lower corruption, greater financial development, and better economic outcomes among other benefits for the society \cite{Johnson:2011aa}.
The concept of trust has also attracted interest in engineering research communities,
ranging from networking to human-machine interaction and artificial intelligence \cite{Ting:2021rc}\cite{Braga:2018qf}\cite{Cho:2015aa}, 
where many problems are cast as buyer-seller interactions \cite{Jung:2019yq}.
The trust game (TG) is a current gold standard of formalisation for non-simultaneous interaction in social dilemma situations 
and has widely been used to study trust and trustworthiness
\cite{Bravo:2008aa}\cite{Johnson:2011aa}\cite{Camerer:1988fk}\cite{Berg:1995aa}\cite{Masuda:2012aa}\cite{McNamara:2009aa}\cite{Tzieropoulos:2013aa}\cite{Lim:2020aa}\cite{Kumar:2020aa}\cite{Capraro:tw}.
The TG is composed of a one-shot sequential interaction between two players in different roles,
one as an investor (representing, for example, a truster, buyer, or citizen) and the other as a trustee (representing, for example, a seller or governor). 
One of the simplest variants of TGs is the binary TG, which involves two strategies per role \cite{Masuda:2012aa}\cite{Guth:2000qq}\cite{dasgupta:2000aa}.
An investor either invests (i.e.\,trusts) or does not invest in a trustee.
Then, the trustee decides to be either trustworthy or untrustworthy to the investor 
(Fig.\,\ref{fig_trust_game}a). 

The evolutionary game theory predicts that
self-interested strategies
(e.g.\,for an investor not to invest) evolve in the two-player binary TG. 
The classical game theory also yields a similar conclusion via backward induction;
given investment from an investor,
a rational trustee is better off by being untrustworthy
and, anticipating it, a rational investor does not invest in a trustee in the first place.
Thus, the two players end up with lower payoffs
than those that they would get with the pro-social strategies 
(i.e.\,for the investor to invest and for the trustee to be trustworthy).
Therefore, an additional mechanism is required for promoting the evolution of the pro-social strategies in the TG~\cite{Manapat:2013aa}\cite{Tarnita:2015aa}\cite{Lim:2022vt}.
	 
An $N$-player binary TG (NTG) was recently proposed 
as a multiplayer (i.e.\,$N\ge 2$) generalisation of the binary TG \cite{Abbass:2016aa}.
However, it suffers from two major difficulties that hamper us from clarifying mechanisms of trust and trustworthiness in multiplayer situations in reasonably realistic manners. First, in this NTG, the investor does not have an option not to invest (Fig.~\ref{fig_trust_game}b); the investor is assumed to invest. Therefore, one cannot investigate the evolution and stability of trusting as opposed to non-trusting behaviour. Note that
their NTG with $N=2$ players is not the two-player TG, which this model attempted to generalise.
Second, investors are allowed to turn into trustees and vice versa by payoff-driven imitation.
An evolutionary outcome of this second assumption is the cease of game playing because all players eventually become trustees \cite{Abbass:2016aa}. Without an investor, one cannot carry on the game. The justification of this result and the underlying assumption of the role-unaware imitation is unclear. The NTG with citizens and governors was used as an example in Ref.~\cite{Abbass:2016aa},
where citizens were allowed to imitate and become governors.
The evolutionary outcome is that all players become governors.
Once there is no citizen, there is no NTG to be played.
A population composed of all governors but no citizen is not only unrealistic but also incompatible with the behavioural experiment setups of the TGs, which ensures that both a citizen (or an investor) and a governor (or a trustee) are always available to play the TG \cite{Johnson:2011aa}.
The follow-up studies of the original NTG \cite{Abbass:2016aa} also inherit the aforementioned two assumptions, i.e., that the investor does not have a choice not to invest and that players can turn into a preferred role by imitation \cite{Chica:2018wt}\cite{Hu:2021tl}\cite{Fang:2021ut}.

In reality, investor-trustee interactions often involve multiplayer interactions rather than dyadic ones;
for instance, multiple investors may be involved in a large project.
Hence, setting up reasonable NTGs and understanding their population dynamics remains a worthwhile goal.
Our contributions in this paper are threefold:
\begin{itemize}
\item We propose an asymmetric NTG with two strategies per role, which generalises the two-player TG but does not suffer from the two problems inherent in the previously proposed NTG.

\item We introduce non-linear payoff functions
that can yield evolutionary dynamics qualitatively different from that of a linear one.

\item We propose an incentive scheme to cost-effectively steer the self-interested players to take prosocial strategies 
such that the population average of the payoff (or social welfare) is maximised. 
\end{itemize}
The source code used for this paper is provided on Github:
\url{https://github.com/iksoolim/asymmetric_N-player_trust_game}.
	 
\section{Model}

\subsection{Population and Group Formation}

We consider an asymmetric NTG in which
the role of each individual is fixed 
as either investor or trustee throughout the whole evolutionary dynamics. Furthermore, we assume that social learning, i.e., payoff-led imitation of strategies, only occurs among individuals of the same role 
as in the two-player TG \cite{Masuda:2012aa}.
There are two strategies available for each role.
An investor either invests or does not invest in trustees.
A trustee selects to be either trustworthy or untrustworthy to investors.
We consider two infinitely large populations, 
one for investors and the other for trustees.
From time to time, a group of $N_I$ investors and  $N_T$ trustees, selected uniformly at random from the respective population, is formed
and these $N \equiv N_I + N_T$ individuals participate in a one-shot NTG. We assume that $N_I$ and $N_T$ are fixed.

\subsection{Payoffs}

We assume that the total value of the investment aggregated over the investing investors is equal to
\begin{equation}
\frac{1-w^{k_i}}{1-w} 
=\begin{cases}
0 & \text{if $k_i =0$},\\
1 & \text{if $k_i =1$}, \\
1+w+w^2+\cdots+w^{k_i-1} & \text{if $k_i \ge 2$},
\end{cases}
\label{eq_gain}
\end{equation}
where $k_i \in \{0,1,\ldots,N_I\}$ denotes  the number of investing investors in the group,
and $w >0$ determines how the value of the investments accumulates 
when an additional investor contributes to the collective good.
A similar non-linear payoff function was previously used for the PGG \cite{hauert2006synergy}.
If $0< w<1$, then the value of the contribution by each additional investing investor is diminishing, i.e.\,discounted or sub-additive.
If $w=1$, then the value of the contribution is $1$ for any investor regardless of the number of investing investors, $k_i$.
This linear payoff function is the same as that for the original NTG~\cite{Abbass:2016aa}.
Note that the total value of the investment is equal to $k_i$ when $w=1$, which follows from L'Hopital's rule applied to the left-hand side of Eq.\,\eqref{eq_gain}.
If $w>1$, the value of the contribution per investor increases as $k_i$ increases, 
i.e.\,representing synergistic or super-additive benefits.

The total investment is equally divided and distributed to the $N_T$ trustees. Therefore, 
the payoff that an untrustworthy trustee in the group receives from the game, denoted by $\Pi_u^o(k_i)$, is given by
\begin{equation}
\Pi_u^o(k_i)  = \frac{1}{N_T} \frac{1-w^{k_i}}{1-w}.
\label{eq_payoff_untrustworthy}
\end{equation}
The payoff of a trustworthy trustee in the group, denoted by $\Pi_t^o(k_i)$, is given by
\begin{equation}
\Pi_t^o(k_i) = r\Pi_u^o(k_i) = r\frac{1}{N_T} \frac{1-w^{k_i}}{1-w},
\end{equation}
where $r$ represents relative productivity of the prosocial strategies and satisfies
$0<r<1$. 
In the two-player TG,
when an investing investor and a trustworthy trustee interact with each other,
each of them gets the same payoff (Fig.\,\ref{fig_trust_game}a).
In the $N$-player generalisation, analogously,
we assume that
when a group of investing investors and a group of trustworthy trustees interact with each other,
each group gets the same (group) payoff.
The aggregated return from the $k_t \in \{0,1,\ldots,N_T\}$ trustworthy trustees
is equally distributed to the $k_i$ investing investors in the group.
Therefore, the payoff that an investing investor receives from the game, denoted by $\Pi_i^o(k_i,k_t)$, is given by
\begin{equation}
\begin{split}
\Pi_i^o(k_i,k_t) 
 & =\underbrace{\frac{1}{k_i} k_t \Pi_t^o(k_i) }_{\text{net gain}} +\underbrace{\left(N_T -k_t\right)\left(-\frac{1}{N_T}\right)}_{\text{net loss}} \\
 & =\frac{k_t}{N_T} \frac{r\left(1-w^{k_i}\right)}{k_i(1-w)} +\left(1 -\frac{k_t}{N_T}\right)\cdot(-1).
\end{split}
\end{equation}
The payoff $\Pi_i^o(k_i,k_t)$ 
is equal to 
the expected payoff of an investing investor playing a two-player game
with each of the $N_T$ trustees; the net gain from a trustworthy trustee is $\frac{r\left(1-w^{k_i}\right)}{k_i(1-w)}$ 
and the net loss from an untrustworthy trustee is $-1$.
Lastly, the payoff of a non-investing investor is $\Pi_n^o  = 0$.
Note that a special case of $N_I =N_T=1$ recovers the two-player TG (Fig.\,\ref{fig_trust_game}a).

By including incentives and associated costs for the players, we define the final payoffs 
$\Pi_i$, $\Pi_n$, $\Pi_t$, and $\Pi_u$ for an investing investor, non-investing investor, trustworthy trustee and untrustworthy trustee, respectively, by
\begin{align}
\Pi_i(k_i,k_t)  &=\Pi_i^o(k_i,k_t) +v_I -a v_I,\\
\Pi_n   & =\Pi_n^o -a v_I,\\
\Pi_t(k_i)  & =  \Pi_t^o(k_i)  +v_T  -a v_T,\\
\Pi_u(k_i)  & =  \Pi_u^o(k_i)  -a v_T,
\end{align}
where an investor pays a fee $a v_I$ to the institution providing the incentives
and an investing investor receives a reward $v_I$ from the institution, where $v_I \ge 0$. 
We assume the fee rate $a>1$ such that the total incentive is less than the total fee,
taking into consideration the operating cost for the institution.
Similarly, a trustee pays a fee $a v_T$ to the institution and a trustworthy trustee receives a reward $v_T \ge 0$.
A similar incentive scheme has been assumed for the PGG \cite{Sasaki:2012rz}.
For a given investor in a group of $N$ players,
the probability that $m_t$ among $N_T$ trustees are trustworthy (and thus $N_T -m_t$ trustees are untrustworthy) is
$\binom{N_T}{m_t} y_t^{m_t}(1-y_t)^{N_T-m_t}$,
where $y_t$ denotes the fraction of trustworthy trustees in the trustee population; $1 -y_t$ is the fraction of untrustworthy trustees.
For a given investor,
the probability that $m_i$ among the other $N_I -1$ investors are investing is
$\binom{N_I -1}{m_i} y_i^{m_i}(1-y_i)^{N_I-1-m_i},
$
where $y_i$ denotes the fraction of investing investors in the investor population.
Therefore, the expected payoff for an investing investor is	 
\begin{equation} 
\begin{aligned}
P_i  & = \sum_{m_i=0} ^{N_I-1} \binom{N_I -1}{m_i} y_i^{m_i}(1-y_i)^{N_I-1-m_i} \\
& \text{\quad} \times \sum_{m_t=0} ^{N_T} \binom{N_T}{m_t} y_t^{m_t}(1-y_t)^{N_T-m_t}\Pi_i(m_i+1,m_t) \\
 & =  \frac{r}{N_I(1 -w)}\frac{y_t }{y_i}\left\{1  -\left[1 +(w-1) y_i\right]^{N_I}\right\}  +y_t    -1 \\
 & \text{\quad} +v_I  -a v_I.
\end{aligned}
\label{eq:P_i}
\end{equation}
Similarly, the expected payoffs $P_n, P_t$ and $P_u$ for a non-investing investor,
trustworthy trustee and untrustworthy trustee, respectively, are given by
\begin{align}
P_n  & =  -a v_I, \label{eq:P_n}\\
P_t &= \sum_{m_i=0} ^{N_I} \binom{N_I}{m_i} y_i^{m_i}(1-y_i)^{N_I-m_i} \notag \\
 & \text{\quad} \times \sum_{m_t=0} ^{N_T-1} \binom{N_T -1}{m_t} y_t^{m_t}(1-y_t)^{N_T-1-m_t}\Pi_t(m_i), \notag \\
 &=  \frac{r}{N_T(1-w)} \left\{1  -\left[1 +(w-1) y_i\right]^{N_I}\right\}    +v_T  -a v_T, \label{eq:P_t}\\
P_u &=  \frac{1}{N_T(1-w)} \left\{1  -\left[1 +(w-1) y_i\right]^{N_I}\right\} - a v_T. \label{eq:P_u}
\end{align}
See Appendix \ref{derivation_payoff} for the derivation Eqs.\,\eqref{eq:P_i}, \eqref{eq:P_t} and \eqref{eq:P_u}.
	
\subsection{Evolutionary Game Dynamics}

For the evolutionary game dynamics,
we use asymmetric replicator equations given by
\begin{align}
\dot{y}_i & =y_i(P_i -P_I) =y_i(1-y_i) (P_i -P_n) \notag \\
& = (1-y_i) y_i \left(\frac{r y_t  \left\{1  -\left[1 +(w-1) y_i\right]^{N_I}\right\} }{N_I (1-w)  y_i} +y_t-1 +v_I\right), \label{eq_replicator1}\\
\dot{y}_t &=y_t(P_t -P_T)=y_t(1-y_t) (P_t -P_u) \notag \\
&= (1-y_t) y_t \left(\frac{(r-1) \left\{1  -\left[1 +(w-1) y_i\right]^{N_I}\right\} }{N_T (1-w)}+v_T\right),
\label{eq_replicator2}
\end{align}
where the dot denotes a time derivative,
$P_I =y_i P_i +(1-y_i) P_n$ is the average payoff of the investor in the entire population, and
$P_T =y_t P_t +(1-y_t) P_u$ is the average payoff of the trustee.
 
To analyse the dynamics given by Eqs.\,\eqref{eq_replicator1} and \eqref{eq_replicator2},
we find all equilibria by setting $\dot{y}_i =\dot{y}_t =0$.
The stability of an equilibrium is determined by the eigenvalues of the Jacobian matrix, which is given by 
\begin{equation}
J
= 
\left(
\begin{array}{cc}
\frac{\partial \dot{y}_i}{\partial y_i} & \frac{\partial \dot{y}_i}{\partial y_t}  \\[5pt]
\frac{\partial \dot{y}_t}{\partial y_i} & \frac{\partial \dot{y}_t}{\partial y_t} 
\end{array}
\right)
= 
\left(
\begin{array}{cc}
J_{11} & J_{12} \\
J_{21} & J_{22} \\
\end{array}
\right)
\label{eq_jacobian}
\end{equation}
at the equilibrium, where
\begin{align}
J_{11} & = r (1-y_i) y_t [(w-1) y_i+1]^{N_I-1} \notag \\
	& \quad -\frac{r y_t \left\{[(w-1)  y_i+1]^{N_I}-1\right\}}{N_I (w-1)} 
	 -(2 y_i-1)   (v_I+y_t-1),
\label{eq:J_11}\\
J_{12} & = \frac{(1-y_i) \left(r \left\{[(w-1)   y_i+1]^{N_I} -1\right\}+N_I (w-1) y_i\right)}{N_I (w-1)},
\label{eq:J_12}\\
J_{21} & = \frac{N_I (r-1) (1-y_t) y_t [(w-1) y_i+1]^{N_I-1}}{N_T},
\label{eq:J_21}\\
J_{22} & =  \frac{(2 y_t-1) (r-1) \left\{1-[(w-1) y_i+1]^{N_I}\right\}}{N_T (w-1)} \notag \\
	& \quad -(2 y_t-1)v_T.
\label{eq:J_22}
\end{align}
If any of the two eigenvalues is positive, the equilibrium is unstable.
Otherwise, the equilibrium is stable;
trajectories starting close enough to the equilibrium remain close enough.
Especially, the equilibrium is asymptotically stable if and only if all the eigenvalues are negative; in this case,
trajectories starting close enough to the equilibrium
converge to it \cite{Strogatz:2015nonlinear}.
Note that Eqs.\,\eqref{eq_replicator1}, \eqref{eq_replicator2} \eqref{eq:J_11}, \eqref{eq:J_12} and~\eqref{eq:J_22} are also valid for $w=1$ with the use of L'Hopital's rule.
		
\section{Results} 

In this section, we characterize the equilibria, their stability, and trajectories of the dynamical system given by
Eqs.\,\eqref{eq_replicator1} and \eqref{eq_replicator2}, of which the state space is $\{ (y_i, y_t) \in [0,1]^2 \}$.
Note that Eqs.\,\eqref{eq_replicator1} and \eqref{eq_replicator2} imply that $(0, 0)$, $(0, 1)$, $(1, 0)$, and $(1, 1)$ are always equilibria.
For proof of the stability of these and the other equilibria, see Appendix \ref{derivation_stability}.

\begin{figure*}[t!]  
\begin{center}  
\includegraphics[width=0.7\textwidth]{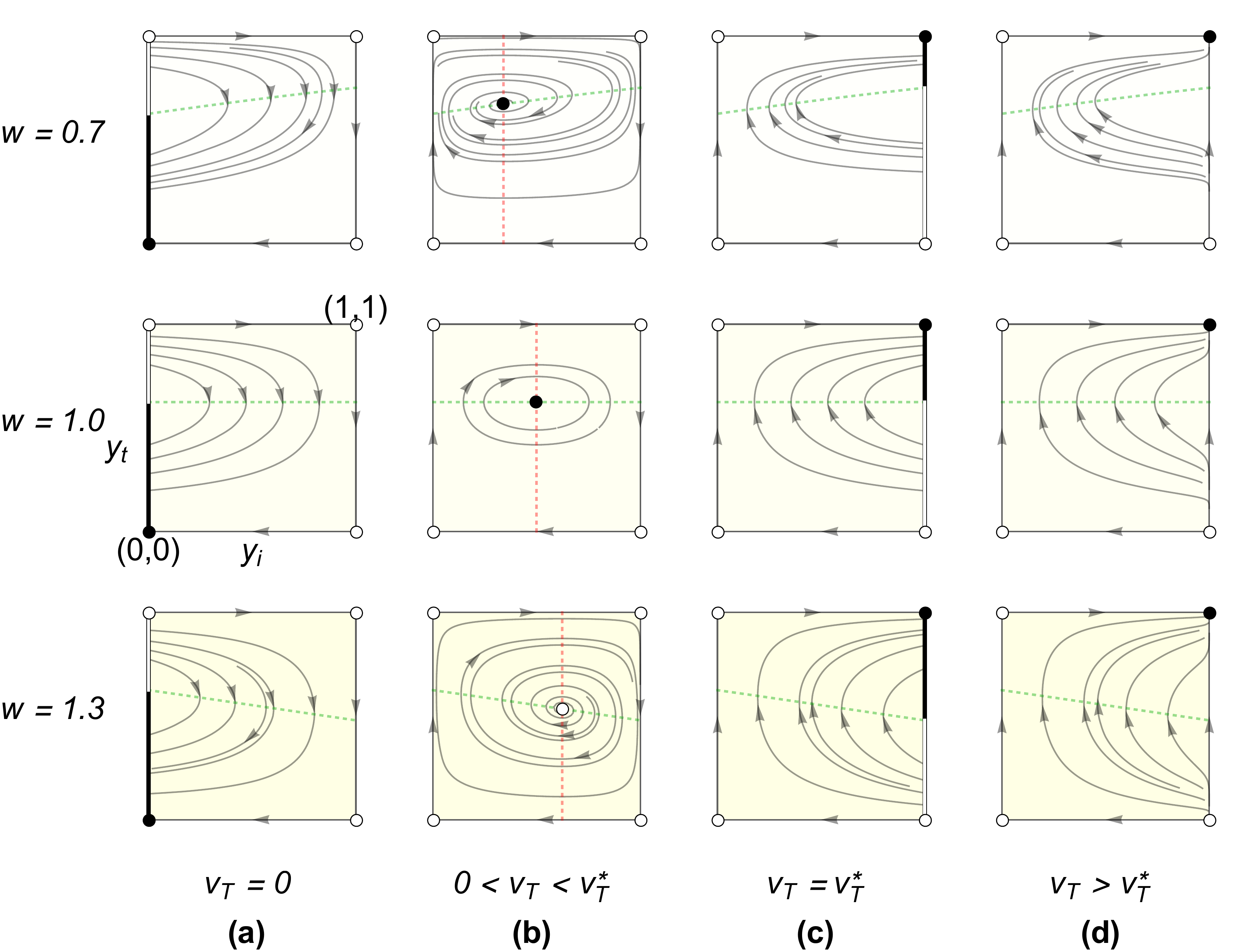}                          
\end{center}  
\caption{ 
Evolutionary game dynamics of the asymmetric NTG with fixed roles for the players.
We set $N_I=5, N_T=5, r=0.6$,
$v_I=0$ and $v_T/v_T^* \in \{0,0.5,1,1.1\}$.
(1st row) $w=0.7$,
(2nd) $w=1$,
and  (3rd) $w=1.3$.
A filled circle represents a stable equilibrium. An open circle represents an unstable equilibrium.
On edges $y_i=0$ and $y_i=1$,
the thick solid lines indicate stable equilibria
and the hollow lines indicate unstable equilibria.
The dashed lines indicate the nullclines $P_i -P_n =0$ (in green) and $P_t -P_u=0$ (in red).
(a) When $v_T =0$ (i.e.\,no incentive to trustworthy trustees),
all trajectories converge to a lower part of the edge $y_i=0$, and investment (i.e.\,trust) does not evolve.
(b) When $0< v_T <v^*_T$,
an interior equilibrium point emerges and moves, with increasing $v_T$, 
from $y_i=0$ towards $y_i=1$.
(c) When $v_T =v^*_T$,
the interior equilibrium disappears
and all trajectories converge to an upper part of the edge $y_i=1$.
(d) When $v_T >v^*_T$,
all trajectories converge to $(1,1)$, i.e., the state of full trust and full trustworthiness.
The nonlinearity in the payoff function yields a stable interior equilibrium with trajectories spiralling into it or an unstable interior equilibrium with trajectories spiralling out of it. These dynamics are qualitatively different from those in the case of the linear payoff function (i.e.\,a neutrally stable interior equilibrium with periodic trajectories around it).
}
\label{fig_dynamics} 
\end{figure*}   

\subsection{$v_T =0$}

For $v_T =0$,
the edge $y_i =0$ of the state space is a line of equilibria.
For $v_T =0 \land 0 \le v_I <1$,
the part of the edge satisfying $0 \le y_t < \frac{1-v_I}{r+1}$, including the origin, $(y_i, y_t) =(0,0)$, is stable but not asymptotically stable (Fig.\,\ref{fig_dynamics}a).
The points on the line satisfying $\frac{1-v_I}{r+1} < y_t \le 1$, including $(0, 1)$, as well as $(1,0)$ and $(1,1)$, are unstable equilibria.
As Fig.\,\ref{fig_dynamics}a indicates, any trajectory is eventually attracted to one of the stable equilibria.
This evolutionary outcome is qualitatively the same as that of the two-player TG and it is so irrespectively of the non-linearity $w$ in the payoff function (e.g.\,for any of $w\in \{0.6, 1, 1.4\}$).
With the special case of $v_T =0 \land v_I =0 \land w=1$, we obtain a baseline model, which is an $N$-player generalisation of the two-player TG
without any other mechanism.

For $v_T =0 \land v_I >1$, the equilibrium $(1,0)$ is not only asymptotically stable but also globally convergent
(i.e.\,reached from any initial state). The equilibria $(0,0)$, $(0, 1)$, $(1,1)$ and $y_i =0$ are unstable.

\subsection{$0< v_T <v^*_T$}

For $0< v_T <v^*_T  \equiv \frac{(1-r) \left(w^{N_I}-1\right)}{N_T (w-1)} \land 0 \le v_I <1$, 
an interior equilibrium
\begin{equation}
\mathbf{Q}
=
\left(
\frac{d^{1/N_I}-1}{w-1},\frac{N_I (1 -v_I)
   \left(d^{1/N_I}-1\right)}{N_I
   \left(d^{1/N_I}-1\right)+(d-1) r}
\right),   
\label{eq:Q-coordinate}   
\end{equation}
emerges, where $d =1 +\frac{N_T v_T(w-1)}{1-r}$.
The interior equilibrium is at the intersection of the two nullclines, $P_i -P_n =0$ and $P_t -P_u=0$ 
with $0 <y_i<1 \land 0<y_t<1$;
see Appendix \ref{existence_interior} for the proof of the existence of the interior equilibrium.
Note that L'Hopital's rule implies that $v^*_T =\frac{N_I (1-r)}{N_T}$ and
$\mathcal{Q} = \left(\frac{N_T v_T}{N_I(1-r)}, \frac{1-v_I}{1+r}\right)$ for $w=1$.

The interior equilibrium is asymptotically stable for $w<1$,
neutrally stable for $w=1$,
and unstable for $w>1$ (Fig.\,\ref{fig_dynamics}b).
The other equilibria are the four corners of the state space, all of which are unstable.	
For $w=1$, at which all the trajectories surrounding $\mathbf{Q}$ form closed cycles,
the time average of $(y_i, y_t)$ over each of the cycles is equal to $(y_i, y_t)$ at $\mathbf{Q}$ given by Eq.\,\eqref{eq:Q-coordinate};
see Appendix \ref{proof_time_average} for the proof.
For $w>1$,  all the trajectories converge to the heteroclinic cycle consisting of the four unstable equilibria, which are saddle points, and the four edges that connect them;
 $(0,0) \rightarrow (0,1)  \rightarrow  (1,1)  \rightarrow (1,0)  \rightarrow (0,0)$.
In this case, the time average of $y_i$ and $y_t$ over the heteroclinic cycle does not converge;
see Appendix \ref{heteroclinic_cycle} for the proof.

For $0< v_T <v^*_T \land v_I >1$, 
there does not exist any interior equilibrium.
In this case, only the four corners are equilibria.
The equilibrium $(1,0)$ is not only asymptotically stable
but also globally convergent. The equilibria $(0,0), (0,1)$ and $(1,1)$ are unstable.

\begin{figure}[t!]  
\begin{center}  
\includegraphics[width=0.45\textwidth]{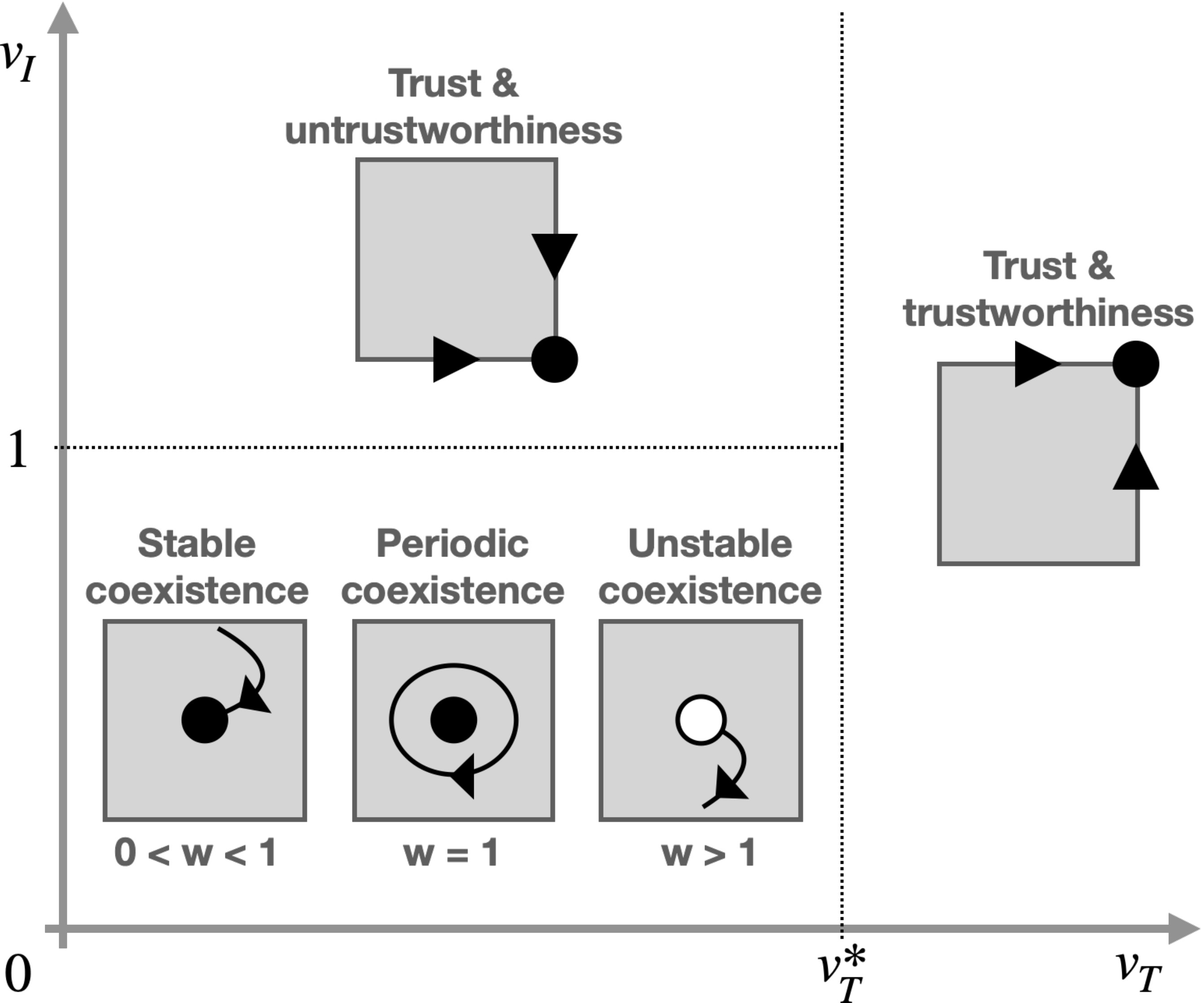}      
\end{center}  
\caption{
Schematic summarising the evolutionary dynamics as a function of the incentive values $v_I$ and $v_T$. On the boundaries of the state space, i.e.\,the unit square, we only show the stable equilibria and trajectories flowing into them. Non-generic cases (i.e.\,$v_T=0, v_T=v^*_T, v_I=0, \text{ and } v_I=1$) are not shown.
}
\label{fig_dynamics_general} 
\end{figure}   

\begin{figure}[t]  
\begin{center}  
\includegraphics[width=0.5\textwidth]{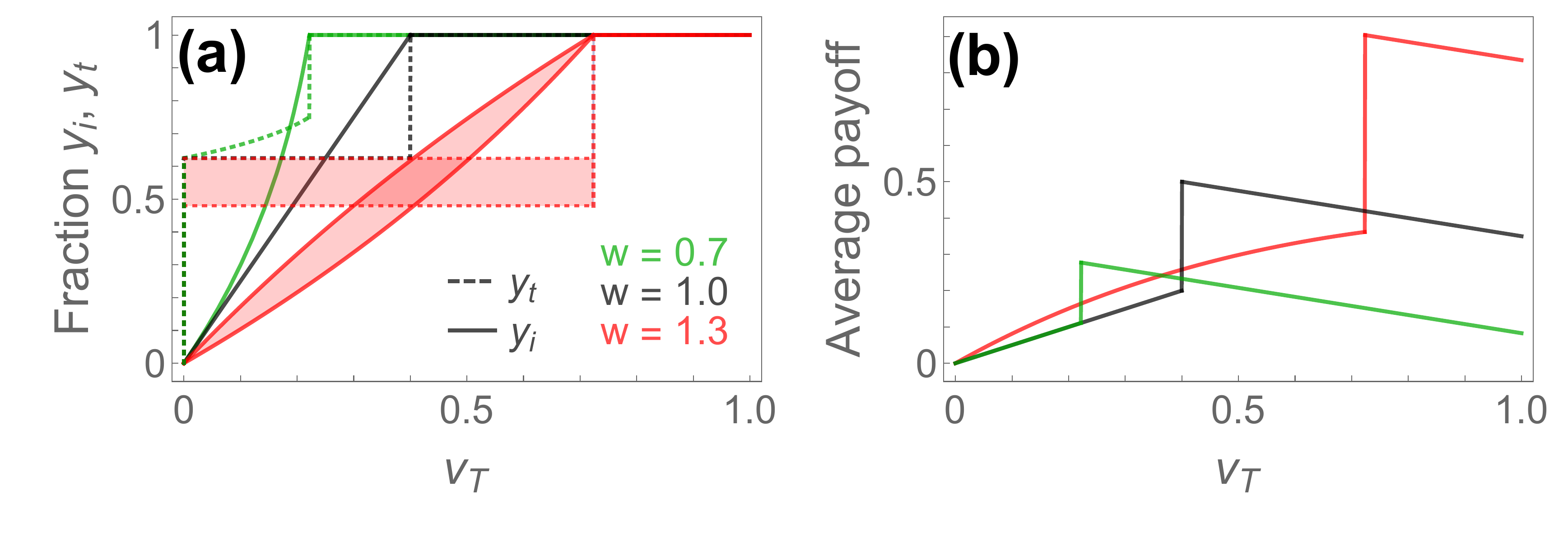}      
\end{center}  
\caption{Effects of the incentive to trustworthy trustees, $v_T$, and the nonlinearity in the payoff function, $w$, on the evolutionary outcomes in the NTG.
We use the same parameter values as those used in Fig.\,\ref{fig_dynamics} except for $v_T$.
(a) Fractions of prosocial players as functions of the reward given to trustworthy trustees, $v_T$, in the equilibrium. We show the fraction of investing investors, $y_i$, and the fraction of trustworthy trustees, $y_t$.
For $w>1$ and $0< v_T <v^*_T$, the time averages of $y_i$ and $y_t$ do not converge. Therefore,
we instead plot the ranges of asymptotic values of $y_i$ and $y_t$ by shaded regions.
We observe that $y_i$ increases as $v_T$  increases when $v_T < v_T^*$.
When $v_T > v_T^*$, the full trust $y_i=1$ and full trustworthiness $y_t=1$ evolve.
(b) Population-averaged payoff, $P$, as a function of $v_T$.
We observe that $P$ increases as $v_T$ increases when $v_T < v_T^*$ and that
$P$ decreases as $v_T$ increases when $v_T > v_T^*$. 
Note that the time average of $P$ converges even if those of $y_i$ and $y_t$ do not.
Panel (a) indicates that as $w$ increases (i.e from sub-linear to linear to super-linear),
the evolution of trust and trustworthiness becomes more difficult. In other words, a higher value of
$v_T$ is necessary for attaining the same fraction of prosocial players when $w$ is larger. In contrast, panel (b) indicates that the payoff of full trust and trustworthiness increases as $w$ increases.}
\label{fig_outcomes} 
\end{figure}   
  
\subsection{$v_T =v^*_T$}

At $v_T =v^*_T$,
a line of equilibria $y_i=1$ emerges.
For $v_T =v^*_T \land 0\le v_I <1$,
the part of the line satisfying $\frac{N_I (1 -v_I) (w-1)}{r (w^{N_I} -1) +N_I (w-1)} < y_t \le 1$, including $(y_i, y_t) = (1, 1)$, is stable but not asymptotically stable (Fig.\,\ref{fig_dynamics}c).
The part of the line satisfying $0 \le \frac{N_I (1 -v_I) (w-1)}{r (w^{N_I} -1) +N_I (w-1)} < y_t$, including $(1, 0)$, and the equilibria $(0,0)$ and $(0,1)$ are unstable.

For $v_T =v^*_T \land  v_I >1$,
the whole line of equilibria including $(1, 0)$ and $(1, 1)$ is stable but not asymptotically stable. The equilibria $(0,0)$ and $(0,1)$ are unstable.
These results are qualitatively the same across the different $w$ values.

\subsection{$v_T >v^*_T$}

For $v_T >v^*_T$, only the four corners are equilibria.
The equilibrium $(1,1)$ is not only asymptotically stable
but also globally convergent (Fig.\,\ref{fig_dynamics}d).
Note that $(1,1)$ represents the fully cooperative populations
entirely consisting of investing investors and trustworthy trustees.
All the other equilibria, namely, $(0,0), (0,1)$ and $(1,0)$, are unstable.
These results hold true  independently of the $v_I  \ge 0$ and $w$ values, except for the dependence of $v^*_T$ on $w$.

In Fig.\,\ref{fig_dynamics_general}, we show a schematic diagram summarising the analysis so far. It presents the evolutionary dynamics that varies in a qualitatively different manner depending on the incentive values $v_I$ and $v_T$.

\subsection{Population Average of Payoff and Optimal Incentive}

One of our goals for proposing and analysing the present NTG is to steer the self-interested players to behave pro-socially, increase the efficiency of the equilibrium in terms of the payoff the players gain and do so in a cost-efficient manner.
Therefore, in this section, we analyze the population average of the payoff given by
\begin{align}
P(y_i,y_t)
=& \frac{N_I}{N_I +N_T} P_I(y_i,y_t)  +\frac{N_T}{N_I +N_T} P_T(y_i,y_t) \notag\\
=& -\frac{a (w-1) (N_I v_I+N_T v_T)+1}{(w-1) (N_I+N_T)} +\frac{N_I (v_I-1)}{N_I+N_T}y_i \notag\\
 &+ \frac{N_T v_T (w-1)-2 r+1}{(w-1)(N_I+N_T)}y_t +\frac{N_I}{N_I+N_T}  y_i y_t \notag\\
 &+ \frac{\left[(2 r-1) y_t+1\right] \left[(w-1) y_i+1\right]^{N_I}}{(w-1) (N_I+N_T)}
\label{eq:P(y_i,y_t)}
\end{align}
after equilibration through the evolutionary dynamics (e.g.\,stable equilibria).
Note $\frac{\partial P}{\partial v_I} = -\frac{N_I (a -y_i)}{N_I+N_T} <0$ since $a>1$ and $y_i \le 1$. In other words, somewhat counterintuitively,
the incentive given to investing investors, $v_I$, harms the overall social welfare
in that the population average of the payoff decreases as $v_I$ increases.
Therefore, for any given $(y_i,y_t)$, one needs to minimise $v_I$ to maximise $P(y_i,y_t)$.

\subsubsection{Optimal Payoff at $(0, 0)$}

The population average of the payoff at $(0,0)$ is given by
\begin{equation}
P(0,0) 
= -\frac{a N_I v_I}{N_I+N_T}.
\end{equation}
If $(0,0)$ is a stable equilibrium (i.e.\,$0\le v_I <1 \land v_T =0$), 
then $P(0,0)$ is maximised at $v_I = 0 \land v_T = 0$.

\subsubsection{Optimal Payoff at $(1, 0)$}

The population average of the payoff at $(1, 0)$ is
\begin{equation}
P(1,0) =\frac{N_I (-a v_I+v_I-1)}{N_I+N_T}+\frac{N_T \left[\frac{1-w^{N_I}}{N_T (1-w)}-a
   v_T\right]}{N_I+N_T}.
\end{equation}
We obtain $\frac{\partial}{\partial v_T} P(1,0)  = -\frac{aN_T}{N_I+N_T} <0$. 
Therefore, if $(1,0)$ is an asymptotically stable equilibrium (i.e.\,$v_I>1 \land 0 \le v_T<v^*_T$),
then $P(1,0)$ is maximised at $v_I=1+\epsilon \land v_T=0$, where $0 < \epsilon \ll 1$.
 
\subsubsection{Optimal Payoff at $(1, 1)$}

The population average of the payoff at $(1,1)$ is
\begin{equation}
P(1,1) 
=\frac{(1-a) (N_I v_I+N_T v_T)}{N_I+N_T} +\frac{2 r \left(w^{N_I}-1\right)}{(w-1) (N_I+N_T)}.
\end{equation}
We obtain $\frac{\partial}{\partial v_T} P(1,1)  =-\frac{(a-1) N_T}{N_I+N_T}<0$.
If $(1,1)$ is an asymptotically stable equilibrium (i.e.\,$v_T >v^*_T$),
then $P(1,1)$ is maximised at $v_I=0 \land v_T=v^*_T +\epsilon$.

\subsubsection{Optimal Payoff at $\mathbf{Q}$ or on Cycles around $\mathbf{Q}$}

Recall that there exists a unique interior equilibrium $\mathbf{Q}$ for $0 \le v_I <1 \land 0<v_T <v^*_T$.
For $w<1$, $\mathbf{Q}$ is an asymptotically stable equilibrium and all the trajectories surrounding $\mathbf{Q}$ converge to it.
For $w=1$, at which all the trajectories surrounding $\mathbf{Q}$ form closed cycles,
the time average of the population-mean payoff over the cycle is the same as the payoff at the equilibrium, i.e., $P(\mathbf{Q})$;
see Appendix \ref{proof_time_average} for the proof.
Therefore, seeking the optimal payoff at $\mathbf{Q}$ is sufficient in both cases $w<1$ and $w=1$.
The population average of the payoff at $\mathbf{Q}$ is given by
\begin{equation}
P(\mathbf{Q})
= \frac{N_T v_T -a (1 -r) (N_I v_I+N_T v_T)}{(1-r)   (N_I+N_T)}.
\end{equation}
Note that $P(\mathbf{Q})$ does not depend on $w$.
We obtain $\frac{\partial P(\mathbf{Q})}{\partial v_T} =\frac{N_T (1-a +a r)}{(1-r)(N_I+N_T)} >0$ when $r >r^*_0 \equiv \frac{a-1}{a}$
and $\frac{\partial P(\mathbf{Q})}{\partial v_T} <0$ when $r <r^*_0$.
Thus, $P(\mathbf{Q})$ is monotonic as a function of $v_T$ (Fig.\,\ref{fig_outcomes}b).
For $0 < w \le 1$,
if $\mathbf{Q}$ is  asymptotically stable (i.e., $w < 1$) or neutrally stable (i.e., $w=1$), then $P(\mathbf{Q})$ is maximised at $v_I =0 \land v_T =v^*_T -\epsilon$ when $r>r^*_0$ 
and at $v_I =0 \land v^*_T =0+\epsilon$ when $r < r^*_0$.

For $w>1$,
the time averages of $y_i$ and $y_t$ do not converge,
but the time average of the payoff converges to 
\begin{equation}
\begin{aligned}
\overline{P}_{\text{hc}} &=\left[\frac{1}{(w-1) \left(\frac{(r+1) (N_I [1-r] + r N_Tv_T)}{N_T v_T \left(r
   \left[w^{N_I}-1\right]+N_I (w-1)\right)}-\frac{r}{w^{N_I}-1}\right)} \right. \\
   & \quad \left. -a (N_I v_I+N_T v_T)\right]\frac{1}{N_I+N_T} ,
\end{aligned}
\end{equation}
where $\overline{P}_{\text{hc}}$ is a convex combination of $P(0,0), P(0,1), P(1,0)$ and $P(1,1)$ as shown in Appendix \ref{heteroclinic_cycle}.
Note that $\frac{\partial \overline{P}_{\text{hc}}}{\partial v_I} =-\frac{a N_I}{N_I+N_T} <0$
and that $\overline{P}_{\text{hc}}$ is monotonic or has a local maximum as a function of $v_T$, as shown in Appendix \ref{optimal_incentive_w>1}.
Therefore, given $v_I = 0$,
the maximum of $\overline{P}_{\text{hc}}(v_T)$
is either $\overline{P}_{\text{hc}}(0 +\epsilon)$, $\overline{P}_{\text{hc}}(v^{\text{hc}}_T)$ or $\overline{P}_{\text{hc}}(v^*_T -\epsilon)$,
where the local maximum of $\overline{P}_{\text{hc}}(v_T)$ is at 
$v_T =v^{\text{hc}}_T \equiv \frac{\left\{\sqrt{aN_I \left(1-r^2\right) (w-1) \left[r \left(w^{N_I}-1\right)+N_I
   (w-1)\right]} -aN_I \left(1-r^2\right) (w-1)\right\}}{a N_T r (w-1) \left(w^{N_I}-N_I w+N_I-1\right)}  \times \left(w^{N_I}-1\right) $.

\subsubsection{Comparison of the Optimal Payoff at the Different Equilibria}  

We now compare the average payoff at the different equilibria. At each equilibrium, including the case of neutral and heteroclinic cycles, we denote by $P^*$ the payoff maximised with respect to $v_I$ and $v_T$. We compare $P^*$ across the different equilibria to seek the overall maximum of the payoff and the associated optimal incentive.

\begin{figure}[t!]  
\begin{center}  
\includegraphics[width=0.48\textwidth]{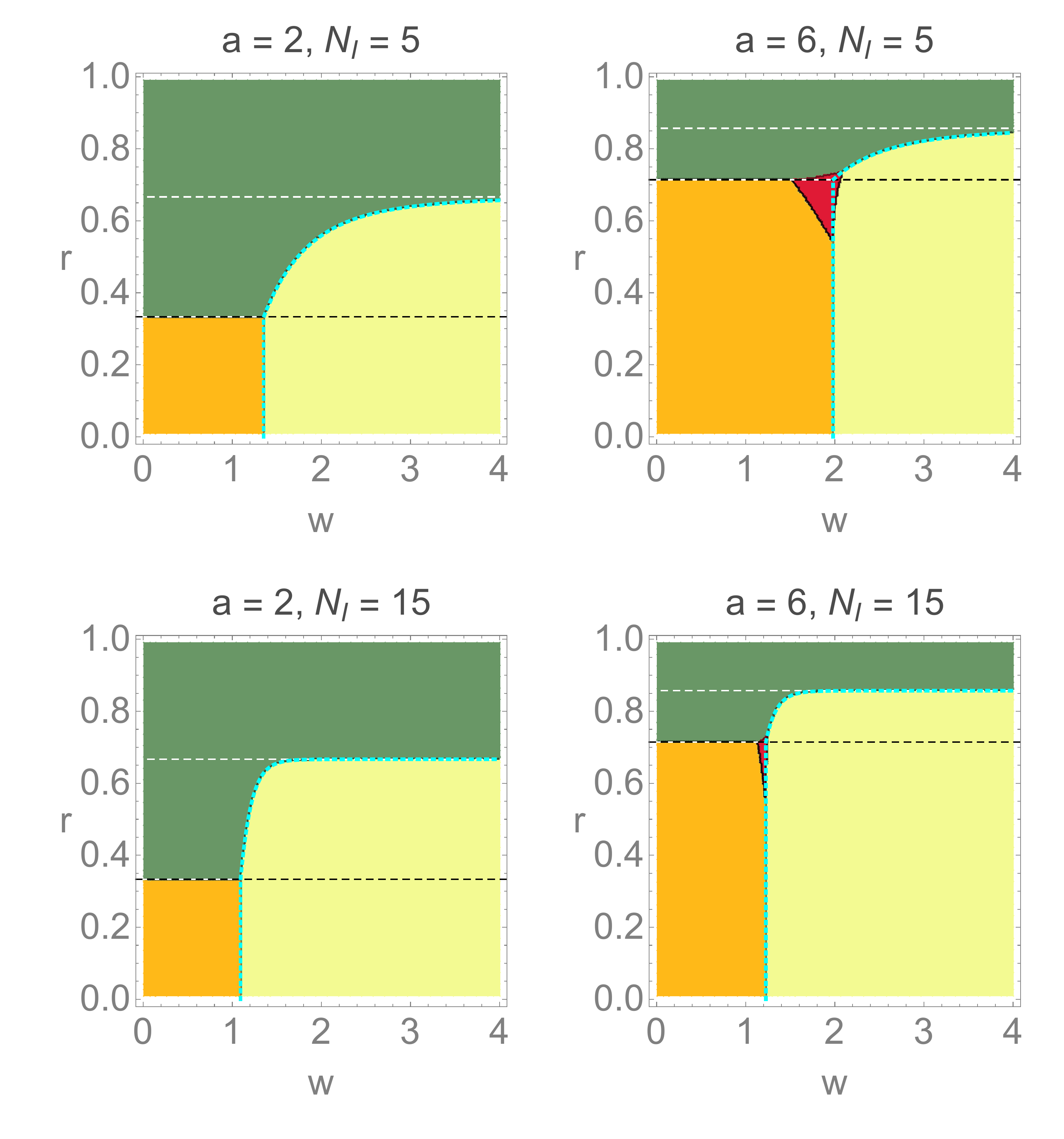}                                                  
\includegraphics[width=0.49\textwidth]{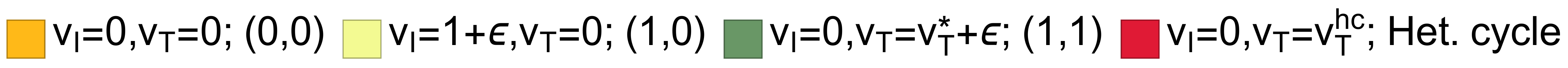}   
\end{center}  
\caption{Optimal incentives and associated evolutionary outcomes. 
Each colored region shows the parameter region in which the associated stable equilibrium or the heteroclinic cycle yields the largest population average of the payoff given by Eq.~\eqref{eq:P(y_i,y_t)}.
Among the two horizontal dashed lines,
the lower and upper ones indicate $r=r^*_1 = \frac{a-1}{a+1} $ and $r=r^*_2 =\frac{a}{a+1}$, respectively.
The dotted curve indicates $r(w)=r^*_2 -\frac{a N_I (w-1)}{(a+1) \left(w^{N_I}-1\right)}$,
where $P^*(1,1) =P^*(1,0)$; note that $r(w) \rightarrow r^*_2$ as $w$ increases.
The vertical dotted line indicates $w=w^*>1$ that we obtained by numerically solving $P^*(0,0) =P^*(1,0)$.
As the fee rate $a$ increases, the parameter region in which $(0, 0)$ is optimal with the null incentive (in orange) and the region in which the heteroclinic cycle is optimal with a positive incentive (in red) become larger.
As $N_I$ or $w$ increases, the border between parameter region in which $P^*(1,1)$ is optimal (in dark green) and that in which $P^*(1,0)$ is optimal (in light yellow) converges to $r=r^*_2$, which we have analytically derived in the limit $N_I \rightarrow \infty$ or $w \rightarrow \infty$.
For a larger fee rate, $a$, or a larger size of the investor group, $N_I$,
the incentive yielding full trust and trustworthiness is optimal for a smaller parameter region (i.e.\,the green regions in the figure).
}
\label{fig_optimal_incentive}
\end{figure}

For $0< w \le 1$,
if $r >r^*_1 \equiv \frac{a-1}{a+1}$, then the optimal payoff among the different equilibria is $P^*(1,1)$;
if $r <r^*_1 $, then the optimal payoff is $P^*(0,0)$;
the associated optimal incentives are $v_I=0 \land v_T=v^*_T +\epsilon$ and $v_I=0 \land v_T=0$, respectively.
For $w > 1$, 
as $N_I \rightarrow \infty$ or $w \rightarrow \infty$,
if $r >r^*_2 \equiv \frac{a}{a+1}$, then the optimal payoff is  $P^*(1,1)$;
if $r <r^*_2$, then  the optimal payoff is  $P^*(1,0)$;
the associated optimal incentives are $v_I=0 \land v_T=v^*_T +\epsilon$ and $v_I=1 +\epsilon \land v_T=0$, respectively.
See Appendix \ref{optimal_incentive} for the derivation of the optimal incentives. 
For relatively small values of $w>1$ and $N_I \ge 2$,
the analytical derivation is not feasible
and we instead numerically obtain the optimal incentives.
Differently from the case of large $N_I$ or $w$,
the incentive yielding the heteroclinic cycle can realize the optimal payoff
(Fig.\,\ref{fig_optimal_incentive}).
Note that copresence of incentives to investors and trustees  (i.e.\,$v_I > 0 \land v_T > 0$) is never optimal.

In summary, if the productivity of the prosocial strategies, $r$, is high enough relative to the fee rate $a$, the incentive leading to the full pro-sociality (i.e.\,full trust and full trustworthiness) is optimal. If the productivity is relatively low, the incentive leading to lower pro-sociality, including the case of the null incentive, is optimal. 
   
\subsection{Other Nonlinear Payoff Functions}

\begin{figure*}[t!]  
\begin{center}  
\includegraphics[width=0.7\textwidth]{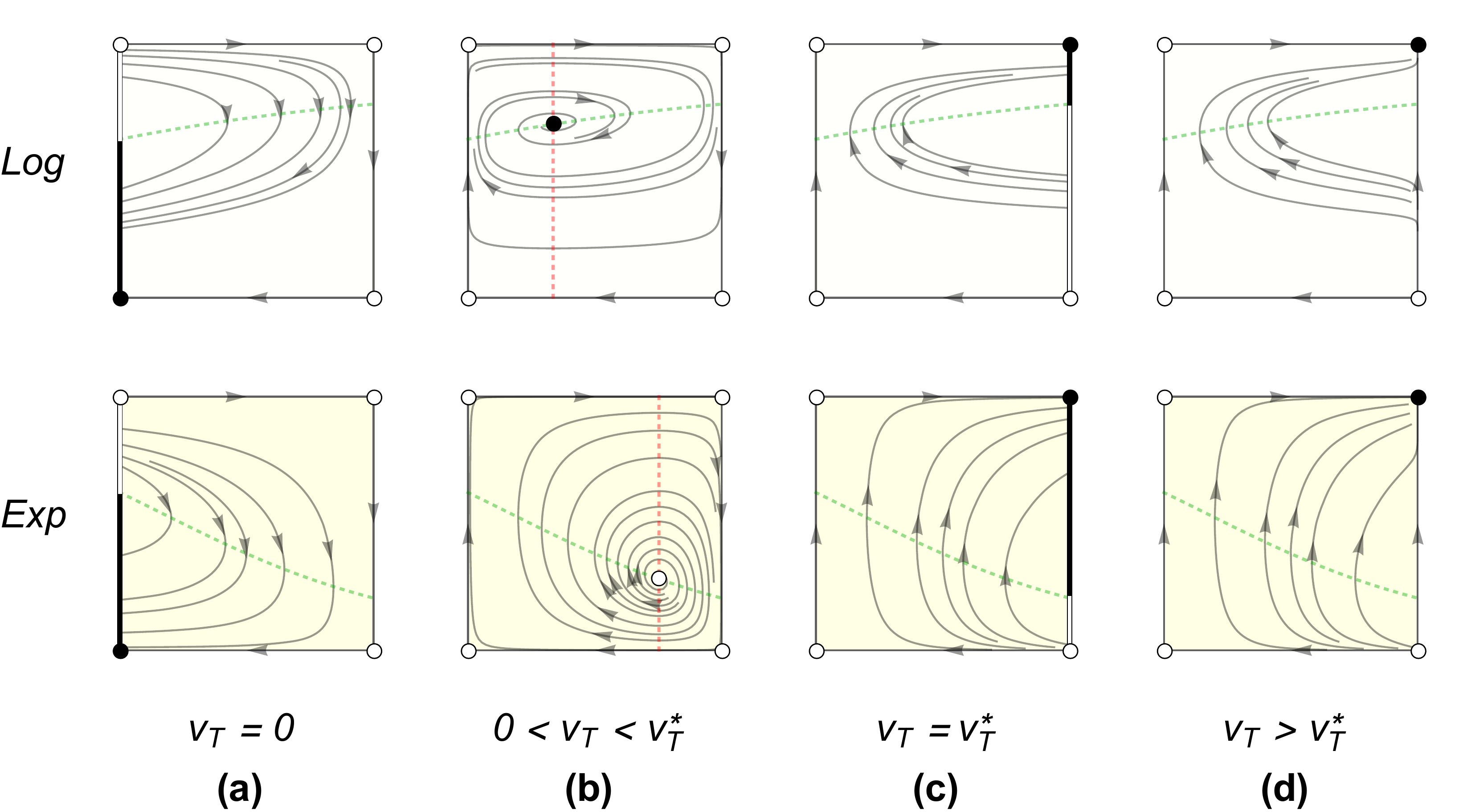}                          
\end{center}  
\caption{ 
Robustness of the evolutionary dynamics with respect to details of nonlinear payoff functions.
The parameters are the same as those used in Fig.\,\ref{fig_dynamics}.
The top panels show that a sub-linear payoff function $\log (k_i +1)/\log (2)$ leads to evolutionary dynamics similar to that for Eq.\,\eqref{eq_gain} with $w=0.7$, which is presented in the top panels in Fig.\,\ref{fig_dynamics}.
The bottom panels show that a super-linear payoff function $\exp(0.7 k_i) -1$ leads to evolutionary dynamics similar to that for Eq.\,\eqref{eq_gain} with $w=1.3$, which is presented in the bottom panels in Fig.\,\ref{fig_dynamics}.
}
\label{fig_robustness}
\end{figure*}

To test the robustness of the results with respect to details of nonlinear payoff functions,
we numerically examine evolutionary dynamics with nonlinear payoff functions that are different from but qualitatively similar to those given by Eq.\,\eqref{eq_gain}. Specifically, we consider $\log (k_i +1)/\log (2)$ as a sub-linear payoff function that is qualitatively similar to Eq.\,\eqref{eq_gain} with $0<w<1$
and $\exp(0.7 k_i) -1$ as a super-linear payoff function that is qualitatively similar to Eq.\,\eqref{eq_gain} with $w>1$.
Figure~\ref{fig_robustness} indicates that each of these payoff functions yields qualitatively the same evolutionary dynamics as those obtained with Eq.\,\eqref{eq_gain}.

\section{Discussion}    
	 
The $N$-player generalisation of a TG game proposed in Ref.~\cite{Abbass:2016aa} assumes that
an investor always invests. Therefore, their NTG is structurally different from both the two-player TG and our NTG.
It may be instead called the trustworthiness game in that the payoff of the game is entirely determined by the strategy of a trustee.
The ultimatum game (UG) and the dictator game (DG)
already have a parallel to this distinction between  the TG  and the trustworthiness game.
The UG involves a non-simultaneous interaction on resource split between a proposer and a responder \cite{Nowak:2000aa}.
The simplest variant of the UG assumes two options for each role:
for a proposer to propose an unfair split in favour of the proposer or a fair split,
and for a responder to accept or reject the proposal.
If the responder accepts, both the proposer and responder obtain the proposed payoffs.
If the response rejects, both players get nothing.
The DG is similar to the UG
except that a responder has no option other than to accept any proposal made by the proposer.
Hence, the payoff entirely depends on what a proposer does and thus the proposer is called a dictator.
The DG is related to but structurally different from the UG, and therefore the DG has been analysed on its own \cite{Guala:2010fk}\cite{Schank:2015wk}\cite{Snellman:2019ta}.
In the UG, the reputation mechanism, which is equivalent to a responder refusing an unfair split, can lead a proposer to offer a fair one \cite{Nowak:2000aa}.
However, the reputation mechanism cannot work for the DG since a responder has no option of refusing any split.
Our NTG is of the UG type in that it allows the investor an option not to invest, which has enabled us to investigate the evolution of trust as well as trustworthiness. 

The evolutionary game dynamics in Ref.~\cite{Abbass:2016aa} assumes role-unaware imitation, which allows imitation between the different roles and leads to the cease of game playing.
The justification of this assumption is unclear.
To the best of our knowledge, this type of game dynamics has not been used prior to Ref.~\cite{Abbass:2016aa}, regardless of two-player or $N$-player games.
In fact, there have been two canonical approaches to modelling evolutionary dynamics of non-simultaneous games.
One approach is to assume that each player plays each role half of the time and imitates others in a role-aware manner \cite{Nowak:2000aa}\cite{Hofbauer:2003aa}\cite{sigmund2016calculus}.
The player's strategy is then a tuple consisting of the strategies under the different roles (e.g.\,one as an investor and the other as a trustee).
This symmetrisation probably better characterises scenarios in which each player has multiple roles, and thus, the payoff of the player is the average of the payoffs from the different roles.
For instance, a bank can lend money to or borrow money from other banks, playing two roles, as a lender/investor and a borrower/trustee.
By this symmetrisation, one can consider the TG using a single population
and the corresponding replicator dynamics \cite{Lim:2022vt}\cite{sigmund2016calculus}.
The same approach has also been used for other asymmetric games 
such as the UG \cite{Nowak:2000aa}\cite{sigmund2016calculus}. Developing and analyzing NTGs with this symmetrisation method is an open question.
A second approach is to fix the two roles such that players can imitate others in their own role only \cite{Masuda:2012aa}\cite{Masuda:2014wz}.
We took this approach to formulate an asymmetric NTG, which is a faithful generalisation of a previously proposed two-player TG \cite{Masuda:2012aa}. Then, differently from the previous work allowing the imitation between the different roles and hence leading to the extinction of investors
\cite{Abbass:2016aa}, we found that investors do not perish but evolve not to trust trustees unless an incentive is in place.

The payoff in Ref.~\cite{Abbass:2016aa} is a linear function of the number of investing investors,
which is also inherited in its follow-up studies \cite{Chica:2018wt}\cite{Hu:2021tl}\cite{Fang:2021ut}.
With a linear payoff function, any $N$-player game is equivalent to a sum of two-player games and thus the evolutionary outcome of the former is similar to that of the latter.
In $N$-player games, however, unlike two-player games, nonlinear payoff functions can yield evolutionary outcomes that are qualitatively different from those of linear ones. We have introduced nonlinear payoff functions in the asymmetric NTG. Even with the nonlinear payoff functions, we have found that it is more challenging for pro-social behaviours to evolve in the asymmetric NTG than in the PGG.
The PGG is one of the most widely studied $N$-player games \cite{archetti2012review}.
With linear payoff functions, the PGG becomes a dominance game
for which anti-social behaviour (i.e.\,defection) dominates pro-social behaviour in terms of the payoff value
and hence only anti-social behaviour evolves.
With the non-linear payoff functions of the same form used in the present paper,
the PGG becomes either a coexistence game or a coordination game
for which prosocial behaviour can evolve \cite{hauert2006synergy}.    
Therefore, incentives have been applied only to the linear PGGs but not the nonlinear PGGs; see Ref.\,\cite{Wang:2021tt} for a review.
In the asymmetric NTG with fixed roles, however, we have found that the nonlinear payoff functions are not sufficient for pro-social behaviour to evolve
and an additional mechanism such as an incentive is required.
We have found that the incentive to trustworthy trustees can be sufficient for the full pro-sociality to evolve in both investor and trustee populations, i.e., the full trust (i.e.\,investment) and the full trustworthiness.
An intuitive explanation of this result is as follows.
If the fraction of trustworthy trustees is high enough,
the payoff of investing investors is higher than that of non-investing ones and thus investing investors evolve.
Hence, if the incentive to trustworthy trustees is large enough for them to evolve, then it also yields the evolution of investing investors.

With the nonlinear payoff function given by Eq.\,\eqref{eq_gain}, one can express the discount (i.e., sub-linear) and synergy (i.e., super-linear) effects by tuning the single parameter $w$. This payoff function is advantageous because it allows us to analytically examine the evolutionary dynamics for arbitrary group sizes $N_I$ and $N_T$. However, our results are not confined to this particular form of payoff function. We ran numerical simulations with different payoff functions to support
that our results are robust with respect to details of the nonlinearity of the payoff function.
We remark that, unlike with Eq.\,\eqref{eq_gain}, different nonlinear payoff functions require separate analyses of evolutionary dynamics for each combination of the values of $N_I$ and $N_T$ in general. Specifically, one needs to numerically find
the interior equilibrium and carry out the linear stability analysis for each given $N_I$ and $N_T$.

Given an investing investor, the two-player TG creates a social dilemma~\cite{Masuda:2012aa}\cite{Abbass:2016aa}.
The total wealth (i.e.\,the sum of the payoffs of an investing investor and a trustee) depends on the strategy of a trustee.
Although a self-interested (i.e.\,untrustworthy) trustee earns higher than a pro-social (i.e.\,trustworthy) trustee does, the former leads to a lower total wealth ($ =0$) than the latter does ($= 2r$) (Fig.\,\ref{fig_trust_game}a). 
Our NTG preserves the nature of a social dilemma. For a linear payoff function,
given the number of investing investors, $k_i$, if all trustees in a group are self-interested, they earn more than any pro-social trustees would.
However, the former leads to a lower total wealth ($= 0$) than the latter does ($= 2r k_i$).

Most previous studies on institutional incentives have focused on which incentives promote prosocial behaviours the best \cite{Perc:2017aa}\cite{Zhang:2020aa}\cite{Fang:2021ut}.
However, a better criterion for the success of an incentive may be the population average of payoff at the evolutionarily stable state \cite{Dong:2019aa}.
Thus, we have sought the optimal incentive that yields the highest payoff, taking into consideration the operating cost of managing incentives.
We have found that the incentive leading to the most prosocial behaviour (i.e.\,full trust and full trustworthiness) often yields the highest payoff but not always.
When the productivity of the prosocial behaviours is not high enough, 
the incentive leading to less prosocial behaviours (e.g.\,combination of full trust and null trustworthiness) can yield the highest payoff;
even the null incentive leading to null trust and null trustworthiness can be optimal 
when the operating cost of managing incentives outweighs benefits from prosocial behaviours.
A limitation of our incentive scheme is to have assumed that an incentive is tailored to individual players while the game is played in groups.
Although this type of the individually targeted incentive is widely used for $N$-player games \cite{Sasaki:2012rz}\cite{Zhang:2020aa}\cite{Cressman:2012tw}\cite{Perc:2012wz},
it may be less feasible than it is for two-player games, in which actions of the individual players are more easily identified than in $N$-player games.
Relaxing this assumption is worthwhile investigation.
For instance, a diluted incentive scheme, which provides an incentive to a group, may be more feasible for $N$-player games. In such an incentive scheme,
all individuals in a group receive the same incentive by construction, and 
whether a group receives an incentive is determined based on aggregated information such as the proportion of trustworthy trustees in the group.

In summary, we started by noting that 
the $N$-player TG in Ref.\,\cite{Abbass:2016aa} is structurally different from the TG and
proposed an asymmetric $N$-player TG with two fixed roles.
With this setup, it is more challenging for pro-social strategies to evolve than in the celebrated PGG. Nonetheless,
we showed that incentives provided to trustees can cost-effectively promote the evolution of trust and trustworthiness among self-interested players. 
We also showed that nonlinear payoff functions in the $N$-player TG yield a richer set of evolutionary dynamics and the associated optimal incentives than linear payoff functions.
We hope that our contribution paves the way for further studies of $N$-player TGs and their variations
such as 
the symmetrisation of asymmetric $N$-player TGs,
the impacts of structured populations \cite{Santos:2008aa}\cite{Alvarez-Rodriguez:2021qy},
repeated interactions on the evolution of trust/trustworthiness,
and stochastic evolutionary dynamics in finite populations.
There can be different generalisations of the two-player NTG each of which recovers the two-player TG when $N_I=N_T=1$;
such generalisations are interesting to explore.
Applications of $N$-player TGs are also worthwhile seeking; 
for instance, multi-hop relay in wireless sensors or ad hoc networks could be mapped to an $N$-player TG among self-interested nodes \cite{Samian:2015td}\cite{Silva:2017aa}.

\appendix
\section{Appendix}
\renewcommand{\theequation}{A.\arabic{equation}}

\subsection{Derivation of Eq.\,\eqref{eq:P_i} \label{derivation_payoff} }

We obtain
\begin{align}
P^o_i &= \sum_{m_i=0} ^{N_I-1} \binom{N_I -1}{m_i} y_i^{m_i}(1-y_i)^{N_I-1-m_i} \notag\\
& \quad \times \sum_{m_t=0} ^{N_T} \binom{N_T}{m_t} y_t^{m_t}(1-y_t)^{N_T-m_t}\Pi^o_i(m_i+1,m_t) \notag\\
&= \sum_{m_i=0} ^{N_I-1} \binom{N_I -1}{m_i} y_i^{m_i}(1-y_i)^{N_I-1-m_i} \notag\\
& \quad \times \sum_{m_t=0} ^{N_T} \binom{N_T}{m_t} y_t^{m_t}(1-y_t)^{N_T-m_t} \left[\frac{m_t}{N_T} \frac{r\left(1-w^{m_i+1}\right)}{(m_i+1)(1-w)} \right. \notag\\
& \left. \qquad\qquad\qquad\qquad\qquad\qquad\qquad\quad +\left(1 -\frac{m_t}{N_T}\right)\cdot(-1)\right] \notag\\
&= \sum_{m_i=0} ^{N_I-1} \binom{N_I -1}{m_i} y_i^{m_i}(1-y_i)^{N_I-1-m_i} \notag\\
& \quad \times \left(\left[\frac{1}{m_i +1} r \frac{1-w^{m_i+1}}{1-w} +1\right] y_t   -1\right) \notag\\
&=  \frac{r y_t}{N_I (1 -w)} \left[\sum_{m_i=0} ^{N_I-1} \binom{N_I}{m_i+1} y_i^{m_i}(1-y_i)^{N_I-1-m_i} \right.	\notag\\ 
& \left. \quad \times(1-w^{m_i+1}) \right]  +y_t  -1 \notag\\
&=  \frac{r}{N_I (1 -w)}\frac{y_t }{y_i}\left[1  -(1 +(w-1)y_i)^{N_I}\right]  +y_t    -1,
\label{eq:P^o_i-derivation}
\end{align}
where we have assumed that $y_i \ne 0$ and used the expression of the mean of a binomial distribution 
$\sum_{m_t=0} ^{N_T} \binom{N_T}{m_t} y_t^{m_t}(1-y_t)^{N_T-m_t} m_t =N_T y_t$ and the relationship
$\binom{N_I -1}{m_i}  \frac{1}{m_i +1} = \frac{1}{N_I}\binom{N_I}{m_i+1}$.
To show the last equality in Eq.~\eqref{eq:P^o_i-derivation}, with substitution $k_i \equiv m_i+1$, we used 
\begin{align}
& \sum_{m_i=0} ^{N_I-1} \binom{N_I}{m_i+1} y_i^{m_i}(1-y_i)^{N_I-1-m_i} \left(1 -w^{m_i+1}\right)\notag\\  
=& \sum_{k_i=1} ^{N_I} \binom{N_I}{k_i} y_i^{k_i-1}(1-y_i)^{N_I-k_i}  \left(1 -w^{k_i}\right) \notag\\
=& \frac{1}{y_i} \sum_{k_i=1} ^{N_I} \binom{N_I}{k_i} y_i^{k_i}(1-y_i)^{N_I-k_i}  \left(1 -w^{k_i}\right) \notag\\
=& \frac{1}{y_i} \left[\sum_{k_i=0} ^{N_I} \binom{N_I}{k_i} y_i^{k_i}(1-y_i)^{N_I-k_i}  \left(1 -w^{k_i}\right) \right. \notag\\
& \left. -\binom{N_I}{0} y_i^0(1-y_i)^{N_I}  \left(1 -w^0\right)\right] \notag\\
=& \frac{1}{y_i}\left[\sum_{k_i=0} ^{N_I} \binom{N_I}{k_i} y_i^{k_i}(1-y_i)^{N_I-k_i}  \left(1 -w^{k_i}\right)\right] \notag\\
=& \frac{1}{y_i} \left[(y_i +1 -y_i)^{N_I} -\sum_{k_i=0} ^{N_I} \binom{N_I}{k_i} y_i^{k_i}(1-y_i)^{N_I-k_i} w^{k_i}\right] \notag\\
=& \frac{1}{y_i} \left[1 -\sum_{k_i=0} ^{N_I} \binom{N_I}{k_i} (w y_i)^{k_i}(1-y_i)^{N_I-k_i}\right] \notag\\
=& \frac{1}{y_i} \left[1 - (w y_i +1 -y_i)^{N_I}\right] \notag\\
=& \frac{1}{y_i} \left[1 - (1 +(w -1)y_i)^{N_I}\right].
\end{align}
We have $P_i =P^o_i +v_I -a v_I$, where $P^o_i$ is given by Eq.~\eqref{eq:P^o_i-derivation}.
We can similarly derive $P_t$ and $P_u$.
	 
\subsection{Existence and Stability of the Equilibria\label{derivation_stability}}

One can deduce the signs of the two eigenvalues $\lambda_1$ and $\lambda_2$ of the Jacobian matrix, $J$, at an equilibrium by its determinant 
and trace, which are equal to $\lambda_1 \lambda_2$ and $\lambda_1 +\lambda_2$, respectively.
We denote by $\text{Det}|_{\mathbf{y}}$ and $\text{Tr}|_{\mathbf{y}}$ the determinant and trace, respectively, of $J$ evaluated at $\mathbf{y} \in [0, 1]^2$. 
Especially, the asymptotical stability of an equilibrium requires $\lambda_1 < 0$ and $\lambda_2 < 0$, which lead to $\text{Det}|_{\mathbf{y}} >0$ and $\text{Tr}|_{\mathbf{y}} <0$.
We determine the stability of each equilibrium as follows.

\subsubsection{$(0,0)$}

The Jacobian matrix at $(y_i, y_t) = (0,0)$ is given by
\begin{equation}
J_{(0,0)} =
\left(
\begin{array}{cc}
 v_I-1 & 0 \\
 0 & v_T \\
\end{array}
\right).
\end{equation}
We obtain
\begin{equation}
\text{Det}|_{(0,0)} = (v_I-1) v_T
\end{equation}
and
\begin{equation}\text{Tr}|_{(0,0)} =v_I+v_T-1.
\end{equation}
If $0 \le v_I < 1  \land v_T =0$,
then $\text{Det}|_{(0,0)} =0 \land \text{Tr}|_{(0,0)} <0$ such that $(0,0)$ is stable but not asymptotically stable.
Otherwise, $(0, 0)$ is unstable.

\subsubsection{$(0,1)$}

The Jacobian at $(0,1)$ is given by
\begin{equation}
J_{(0,1)} =
\left(
\begin{array}{cc}
r+v_I & 0 \\
0 & -v_T \\
\end{array}
\right).
\end{equation}

We obtain
\begin{equation}
\text{Det}|_{(0,1)}= -v_T (r +v_I)
\end{equation}
and
\begin{equation}\text{Tr}|_{(0,1)} = r +v_I-v_T.
\end{equation}
If $v_T =0$, then $\text{Det}|_{(0,1)} =0 \land \text{Tr}|_{(0,1)} >0$ such that $(0,1)$ is unstable.
If $v_T >0$, then $\text{Det}|_{(0,1)} <0$ such that $(0,1)$ is unstable.

\subsubsection{$(1,0)$}

The Jacobian  at $(1,0)$ is given by
\begin{equation}
\begin{aligned}
J_{(1,0)} 
&=\left(
\begin{array}{cc}
 1-v_I & 0 \\
 0 & v_T -\frac{(1-r) \left(w^{N_I}-1\right)}{N_T (w-1)}\\
\end{array}
\right) \\
&=\left(
\begin{array}{cc}
 1-v_I & 0 \\
 0 & v_T -v^*_T\\
\end{array}
\right).
\end{aligned}
\end{equation}
We obtain
\begin{equation}
\text{Det}|_{(1,0)} =\left(v_T-v_T^*\right) (1-v_I)
\end{equation}
and
\begin{equation}
\text{Tr}|_{(0,1)} =v_T -v_T^* +1 -v_I,
\end{equation}
where
\begin{equation}
v_T^* =\frac{(1-r) \left(w^{N_I}-1\right)}{N_T (w-1)} >0.
\end{equation}
If $v_I >1 \land v_T< v_T^*  $, then $\text{Det}|_{(1,0)} >0 \land \text{Tr}|_{(1,0)} <0$ such that $(1,0)$ is asymptotically stable.
If $\left(v_I > 1 \land v_T= v_T^*\right) \lor \left(v_I = 1 \land v_T <v_T^*\right)$, then $\text{Det}|_{(1,0)} =0 \land \text{Tr}|_{(1,0)} <0$ such that $(1,0)$ is stable but not asymptotically stable.
Otherwise, $(1, 0)$ is unstable.
     
\subsubsection{$(1,1)$}

The Jacobian  at $(1,1)$ is given by
\begin{equation}
\begin{aligned}
J_{(1,1)} 
&=\left(
\begin{array}{cc}
 -\frac{r \left(w^{N_I}-1\right)}{N_I (w-1)}-v_I & 0 \\
 0 & \frac{(1-r)(w^{N_I} -1)}{N_T (w-1)} -v_T\\
\end{array}
\right) \\
&=\left(
\begin{array}{cc}
 -\frac{N_T r   v_T^*}{N_I (1-r)} -v_I & 0 \\
 0 &  v^*_T -v_T\\
\end{array}
\right).
\end{aligned}
\end{equation}
We obtain
\begin{equation}
\text{Det}|_{(1,1)} =\left(v_T-v_T^*\right) \left[v_I +\frac{N_T r   v_T^*}{N_I (1-r)}\right]
\end{equation} 
and
\begin{equation}
\text{Tr}|_{(1,1)} =\left[1 -\frac{N_T r}{N_I (1 -r)}\right] v_T^* -v_I-v_T.
\end{equation}
We obtain $\text{sign}(\text{Det}|_{(1,1)}) =\text{sign}\left(v_T-v_T^*\right)$
since $v_I +\frac{N_T r   v_T^*}{N_I (1-r)} >0$, which is guaranteed by $0<r<1$, $v_I \ge 0$ and $v^*_T >0$.
If $v_T >\left[1 -\frac{N_T r}{N_I (1 -r)}\right] v_T^* -v_I$,
then $\text{Tr}|_{(1,1)} <0$.
We also note $\left[1 -\frac{N_T r}{N_I (1 -r)}\right] v_T^* -v_I \le \left[1 -\frac{N_T r}{N_I (1 -r)}\right] v_T^* < v_T^*$.
Therefore, if $v_T> v_T^*$, then $\text{Det}|_{(1,1)} >0  \land \text{Tr}|_{(1,1)} <0$ such that $(1, 1)$ is asymptotically stable.

If $v_T = v_T^*$, then $\text{Det}|_{(1,1)} =0  \land \text{Tr}|_{(1,1)} <0$. Therefore, $(1, 1)$ is stable but not asymptotically stable.

If $v_T< v_T^*$, then $\text{Det}|_{(1,1)} <0$. Therefore, $(1, 1)$ is unstable.

\subsubsection{Interior equilibrium $\mathbf{Q}$ \label{existence_interior}}

We show that there exists a unique interior equilibrium $\mathbf{Q}$
if and only if $0< v_T < v^*_T =\frac{(1-r) \left(w^{N_I}-1\right)}{N_T (w-1)}$ and $v_I <1$.
The internal equilibrium, if it exists, is located at the intersection of the nullclines $P_t(y_i,y_t) -P_u(y_i,y_t) =0$ and $P_i(y_i,y_t) -P_n(y_i,y_t) =0$ with $0<y_i<1 \land 0<y_t<1$. Let us investigate the two nullclines one by one.

Because $P_t -P_u =\frac{(1-r) \left\{1-[(w-1) y_i+1]^{N_I}\right\}}{N_T (w-1)}+v_T$ does not depend on $y_t$,
the nullcline $P_t -P_u=0$ is of the form $y_i =\text{ constant}$. Specifically, $P_t -P_u=0$ leads to $y_i =y_{i,Q} \equiv \frac{d^{1/N_I}-1}{w-1}$,
where $d =1 +\frac{N_T v_T(w-1)}{1-r}$.
We obtain $\frac{d}{d y_i} \left[P_t -P_u\right] = -\frac{N_I (1-r) [(w-1) y_i+1]^{N_I-1}}{N_T} <0$.
Therefore,
if and only if $0< v_T < v^*_T$,
then $P_t(0, y_t) -P_u(0, y_t) =v_T >0$ and $P_t(1, y_t) -P_u(1, y_t) = v_T -v^*_T <0$
such that the nullcline $P_t -P_u =0$ (i.e.\,$y_i =y_{i,Q}$)  exists with $0<y_{i, Q} <1$.

To examine the other nullcline, we look into $ P_i -P_n = \frac{r y_t  \left\{1  -\left[1 +(w-1) y_i\right]^{N_I}\right\} }{N_I (1-w)  y_i} +y_t-1 +v_I$.
In fact, $\frac{\partial}{\partial y_t} \left[P_i -P_n\right] >0$ and $P_i(y_i,1) -P_n(y_i,1) >0$ hold true for $0<y_i<1$, which we will show later.
Therefore, if and only if $v_I <1$, then $P_i(y_i,0) -P_n(y_i,0) =v_I-1 <0$
such that the nullcline $P_i -P_n =0$  exists in the range $0<y_t<1$.
Note that the nullcline $P_i -P_n =0$ can be represented by $y_t =g(y_i)$
because there exists a unique $y_t$ satisfying $P_i -P_n =0$ for any $y_i$.

We now show $\frac{\partial}{\partial y_t} \left[P_i -P_n\right] =\frac{r  \left\{1  -\left[1 +(w-1) y_i\right]^{N_I}\right\} }{N_I (1-w)  y_i} +1 >0$ for $0<y_i <1$.
If $0<w<1$, then we obtain $0< 1 +(w-1) y_i <1$ such that $1  -\left[1 +(w-1) y_i\right]^{N_I}$ and $1-w$  are both positive.
If $w >1$, then $1 < 1 +(w-1) y_i$ such that $1  -\left[1 +(w-1) y_i\right]^{N_I}$ and $1-w$  are both negative.
If $w=1$, then $\frac{\partial}{\partial y_t} \left[P_i -P_n\right] =\frac{\lim_{w\rightarrow 1} r  \left\{1  -\left[1 +(w-1) y_i\right]^{N_I}\right\} }{\lim_{w\rightarrow 1} N_I (1-w)  y_i} +1 =r +1  >0$. Therefore, we have proved $\frac{\partial}{\partial y_t} \left[P_i -P_n\right] > 0$ for any $w$.

We now show $P_i(y_i,1) -P_n(y_i,1) >0$.
If $w\ne 1$,
then we obtain $P_i(y_i,1) -P_n(y_i,1) =\frac{r  \left\{1  -\left[1 +(w-1) y_i\right]^{N_I}\right\}}{N_I (1-w)y_i}+v_I >0$.
If $w=1$,
then we obtain $P_i(y_i,1) -P_n(y_i,1) 
=\frac{\lim_{w\rightarrow 1} r  \left\{1  -\left[1 +(w-1) y_i\right]^{N_I}\right\} }{\lim_{w\rightarrow 1} N_I (1-w)  y_i} +v_I
=r +v_I > 0$. Therefore, $P_i(y_i,1) -P_n(y_i,1) >0$ holds true for any $w$.

Finally, these results imply that there is a unique intersection of $y_i =y_{i,Q}$ and $y_t =g(y_i)$ satisfying $0<y_i<1 \land 0<y_t<1$, which is an interior equilibrium $\mathbf{Q}$.

We now analyse the stability of the interior equilibrium $\mathbf{Q}$.	
The Jacobian  at $\mathbf{Q}$ is given by
\begin{equation}
J_Q =
\left(
\begin{array}{cc}
J^Q_{11} & J^Q_{12} \\[5pt]
J^Q_{21} & 0 \\
\end{array}
\right),
\end{equation}
where
$J^Q_{11} = \frac{N_I w d - d^{1/N_I} \left\{d [(N_I-1) w+N_I]+w\right\}+[d(N_I-1)+1] d^{2/N_I}}
{(w-1) \left\{d^{1/N_I} [N_I-(d-1) r]-N_I d^{2/N_I}\right\}} \times r (1-v_I)$,
$J^Q_{12} = - \frac{\left\{N_I d^{1/N_I} + [(d-1) r-N_I]\right\}  \left(d^{1/N_I}-w\right)}{N_I (w-1)^2}$,
and
$J^Q_{21}  =  -\frac{d^{1-1/N_I} \left(d^{1/N_I}-1\right) \left\{N_I v_I d^{1/N_I}+
  [(d-1) r-N_I v_I]\right\}} {N_T \left\{N_I d^{1/N_I}+[(d-1) r-N_I]\right\}^2} \times N_I^2 (r-1) (v_I-1)$.
We first show $\text{Det}|_Q > 0$ and $\text{sign}(\text{Tr}|_Q ) =\text{sign}(w-1)$.

For $w\ne 1$, we have
$\text{Det}|_Q 
= (v_I -1)N_I (1-r)  d^{1-1/N_I} 
 \frac{
   \left(d^{1/N_I}-1\right) \left(d^{1/N_I}-w\right)
   \left[N_I v_I \left(d^{1/N_I}-1\right)+(d-1) r\right]}{N_T
   (w-1)^2 \left[N_I \left(d^{1/N_I}-1\right)+(d-1) r\right]}
$.
We note that $\frac{N_I v_I \left(d^{1/N_I}-1\right)+(d-1) r}{N_I \left(d^{1/N_I}-1\right)+(d-1) r}$ is positive because $\text{sign}(d-1) =\text{sign}\left(d^{1/N_I} -1\right)$.
Since $0<v_T <v_T^* =\frac{(1-r) \left(w^{N_I}-1\right)}{N_T (w-1)}$ 
for the existence of the interior equilibrium,
we have $d=1 +\frac{N_T v_T (w-1)}{1-r} 
= 1 +s \left(w^{N_I}-1\right)$,
where $v_T =s v^*_T$ and $0< s <1$.
Therefore, we obtain
$w^{N_I} -d =(1-s) \left(w^{N_I}-1\right) \implies \text{sign}\left(w^{N_I} -d\right) =\text{sign}\left(w^{N_I} -1\right) \implies \text{sign}\left(w -d^{1/N_I}\right) =\text{sign}\left(w -1\right) \implies \left(w <d^{1/N_I} <1\right) \lor \left(1 <d^{1/N_I} <w\right)  \implies \left(d^{1/N_I}-1\right) \left(d^{1/N_I}-w\right) <0 \implies \text{sign}\left(\left(d^{1/N_I}-1\right) \left(d^{1/N_I}-w\right)\right)= -1$.
For $ d\ne 1$,
we obtain $\text{sign} \left(\text{Det}|_Q \right) =\text{sign} \left(1 -v_I \right)$
because 
$\text{sign} \left(\text{Det}|_Q\right)
=\text{sign} \left(v_I-1\right)         \text{sign} \left[\left(d^{1/N_I}-1\right) \left(d^{1/N_I}-w\right)\right]
	\times \text{sign} \left(\frac{N_I v_I \left(d^{1/N_I}-1\right)+(d-1) r}{N_I \left(d^{1/N_I}-1\right)+(d-1) r} \right)
= (-1)\cdot (-1)\cdot 1
=1
$.
Recall that $0\le v_I <1$ is required for the existence of $\mathbf{Q}$.
For $w=1$, we obtain $\text{Det} |_Q = \frac{(1-v_I) v_T (r+v_I) [N_I (1-r)-N_T v_T]}{N_I
   \left(1-r^2\right)}>0$
since  $0<v_T<v_T^* =\frac{\lim_{w\rightarrow 1} (1-r) \left(w^{N_I}-1\right)}{\lim_{w\rightarrow 1}N_T (w-1)}  =\frac{N_I (1-r)}{N_T}$ is required for the existence of $\mathbf{Q}$.
Hence, we have shown $\text{Det} |_Q >0$ or $\text{sign}(\text{Det} |_Q) =1$ regardless of the $w$ value.
 		
For $w\ne 1$,
we obtain
$\text{Tr}|_Q  =
\frac{r (1-v_I) d^{-1/N_I} \left\{ d^{1/N_I} \left[(d (N_I-1)+1\right]-d  N_I\right\}}{N_I
   \left(d^{1/N_I}-1\right)+(d-1) r} \frac{w -d^{1/N_I}}{w-1}$.	
We obtain $\frac{\text{sign} \left(w -d^{1/N_I}\right)}{\text{sign}(w-1)} =1$ since $\text{sign} \left(w -d^{1/N_I}\right) = \text{sign}(w-1)$ as already shown.
For $d\ne 1$, we have $q(d) \equiv  d^{1/N_I} \left[d (N_I-1)+1\right] - d N_I >0$
since $q(1) = 0$ is the global minimum of $q(d)$ for $d>0$,
the latter of which can be shown as follows.
First, $q(1)$ is a local minimum of $q(d)$ since $\left. \frac{\partial q}{\partial d} \right|_{d=1} = \left. \left\{\frac{d^{1/N_I -1} \left[d\left(N_I^2-1\right)+1\right]}{N_I}-N_I\right\} \right|_{d=1} =0$
and $\left. \frac{\partial^2 q}{\partial d^2} \right|_{d=1} = \left. \frac{(N_I-1) d^{1/N_I -2} (dN_I+d-1)}{N_I^2} \right|_{d=1}  =\frac{N_I-1}{N_I} >0$.
Second, $d=d^* \equiv \frac{1}{N_I +1} \in (0,1)$ is the only inflection point of the function $q(d)$ for $d>0$ 
since $\frac{\partial^2 q}{\partial d^2}<0$ for $d<d^*$, $\frac{\partial^2 q}{\partial d^2} =0$ at $d=d^*$, and $\frac{\partial^2 q}{\partial d^2} >0$ for $d>d^*$.
Therefore, there is no local minimum in $d \le d^*$ and at most one local minimum in $d > d^*$, which is at $d=1$.
Third, we obtain $q(0)=0 = q(1)$. Hence, $q(1)=0$ is the global minimum of $q(d)$ for $d>0$.
We obtain $\text{sign}(N_I\left(d^{1/N_I}-1\right)+(d-1) r) =\text{sign}(w-1)$
since $ \text{sign}(d^{1/N_I} -1) =\text{sign}(d-1) = \text{sign}(w -1)$.        
It follows that $\text{sign}(\text{Tr}|_Q ) 
=\text{sign}(1-v_I)\frac{\text{sign}\left(d^{1/N_I}\left[(d (N_I-1)+1\right]-d  N_I\right)}{\text{sign}\left(N_I
   \left(d^{1/N_I}-1\right)+(d-1) r\right)} \frac{\text{sign} \left(w -d^{1/N_I}\right)}{\text{sign}(w-1)}
=1 \cdot \frac{1}{\text{sign}(w-1)} \cdot 1
=\text{sign}(w-1)
$.
For $w=1$, it holds true that $\text{sign}(\text{Tr}|_Q ) =0$ because $\text{Tr}|_Q = 0$.
Hence, we have shown $\text{sign}(\text{Tr}|_Q ) =\text{sign}(w-1)$ regardless of the $w$ value.

For $w<1$, we obtain $\text{Det}|_Q > 0 \land \text{Tr}|_Q <0$ such that $\mathbf{Q}$ is asymptotically stable.
For $w >1$, we obtain $\text{Det}|_Q > 0 \land \text{Tr}|_Q >0$ such that $\mathbf{Q}$ is unstable. 
For $w=1$, we obtain $\text{Det} |_Q >0$ and $\text{Tr} |_Q =0$.
In this case, the discriminant $D=\left(\text{Tr}|_Q\right)^2 -4\,\text{Det}|_Q <0$ and $\text{Tr} |_Q =0$, which implies that
the eigenvalues are purely imaginary. Therefore, $\mathbf{Q}$ is neutrally stable
and the trajectories cycle around it.

\subsubsection{$y_i =0$}

We find that $(0,y_t)$, where $0< y_t <1$, is a line of equilibria if and only if $v_T = 0$.
In this case, the Jacobian at $(0,y_t)$ is given by
\begin{equation}
J_{(0,y_t)} =
\left(
\begin{array}{cc}
 r y_t+v_I+y_t-1 & 0 \\
 -\frac{N_I (r-1) (y_t-1) y_t}{N_T} & 0 \\
\end{array}
\right).
\end{equation}

We obtain $\text{Det}|_{(0,y_t)} =0$ and $\text{Tr}|_{(0,y_t)} =   (r+1) y_t +v_I -1$.
If $v_T=0 \land y_t <\frac{1-v_I}{r+1}$, then $\text{Tr}|_{(0,y_t)}<0$ such that $(0, y_t)$ is stable but not asymptotically stable.
If $v_T=0 \land y_t >\frac{1-v_I}{r+1}$, then $\text{Tr}|_{(0,y_t)} >0$ such that $(0, y_t)$ is unstable.

\subsubsection{$y_i =1$}

We find that $(1,y_t)$, where $0< y_t <1$, is a line of equilibria if and only if $v_T=v^*_T$. 
In this case,
the Jacobian at $(1,y_t)$ is given by
\begin{equation}
J_{(1,y_t)} =
\left(
\begin{array}{cc}
 -\frac{r y_t \left(w^{N_I}-1\right)}{N_I (w-1)}-v_I-y_t+1 & 0 \\
 -\frac{N_I (r-1) (y_t-1) y_t w^{N_I-1}}{N_T} & 0 \\
\end{array}
\right).
\end{equation}
We obtain $\text{Det}|_{(1,y_t)} =0$ and $\text{Tr}|_{(1,y_t)} = -\frac{r y_t \left(w^{N_I}-1\right)}{N_I (w-1)} -v_I-y_t+1$.
If $v_T=v^*_T \land 0\le v_I <1 \land y_t >\frac{N_I (1 -v_I) (w-1)}{r (w^{N_I} -1) +N_I (w-1)}$, then $\text{Tr}|_{(0,y_t)} <0$ such that $(1, y_t)$ is stable but not asymptotically stable,
where $0 <\frac{N_I (1 -v_I) (w-1)}{r (w^{N_I} -1) +N_I (w-1)} <1$.
If $v_T=v^*_T \land 0\le v_I <1 \land y_t <\frac{N_I (1 -v_I) (w-1)}{r (w^{N_I} -1) +N_I (w-1)}$, then $\text{Tr}|_{(0,y_t)} >0$ such that $(1, y_t)$ is unstable.
If $v_T=v^*_T \land v_I >1$, then $\text{Tr}|_{(0,y_t)} <0$ such that the entire line of equilibria is stable.

\subsubsection{$y_t =0$}

We find that $(y_i,0)$, where $0< y_i <1$, is a line of equilibria if and only if $v_I =1$. 
In this case,
the Jacobian at $(y_i,0)$ is given by
\begin{equation}
J_{(y_i,0)} =
\left(
\begin{array}{cc}
 0 & -\frac{(y_i-1) \left(r \left(((w-1) y_i+1)^{N_I}-1\right)+N_I
   (w-1) y_i\right)}{N_I (w-1)} \\
 0 & \frac{(r-1)\left( [(w-1) y_i+1]^{N_I}-1\right)}{N_T (w-1)} + v_T \\
\end{array}
\right).
\end{equation}
We obtain 
\begin{eqnarray}
\text{Det}|_{(y_i,0)} &=& 0,\\
\text{Tr}|_{(y_i,0)} &=& \frac{(r-1)\left( [(w-1) y_i+1]^{N_I}-1\right)}{N_T (w-1)} + v_T \notag\\
\end{eqnarray}
and 
\begin{equation}
\frac{\partial}{\partial y_i} \text{Tr}|_{(y_i,0)}  
= -\frac{N_I (1-r) [1+(w-1) y_i]^{N_I-1}}{N_T} <0.
\end{equation}
If $v_T =0$, then $\text{Tr}|_{(y_i,0)}>0$  such that $(y_i,0)$ is unstable.
If $0<v_T <v^*_T \land y_i > \frac{\left(\frac{N_T  v_T(1-w)}{r-1} +1\right)^{1/N_I}-1}{w-1} =\frac{d^{1/N_I}-1}{w-1}$,
then $\text{Tr}|_{(y_i,0)} <0$ such that $(y_i,0)$ is stable but not asymptotically stable.
If $0<v_T <v^*_T \land y_i < \frac{d^{1/N_I}-1}{w-1}$,
then $\text{Tr}|_{(y_i,0)} >0$ such that $(y_i,0)$ is unstable.
Note that we obtain $0 <\frac{d^{1/N_I}-1}{w-1} <1$ for $0<v_T <v^*_T$.
If $v_T \ge v^*_T$,
then $\text{Tr}|_{(y_i,0)}>0$ such that $(y_i,0)$ is unstable.

\subsubsection{$y_t =1$}

There is no equilibrium on the edge $(y_i,1)$. This is because
$\dot{y_i}  = (1-y_i) y_i \left(\frac{r  \left\{1  -\left[1 +(w-1) y_i\right]^{N_I}\right\} }{N_I (1-w)  y_i} +v_I\right) 
>0$, which follows from the combination of
$\frac{1  -\left[1 +(w-1) y_i\right]^{N_I}}{1-w} >0$ shown in Appendix \ref{existence_interior} and
$0<y_i <1$.
 	
\subsection{Time Average of $(y_i, y_t)$ and the Payoff over a Cycle for $w=1$\label{proof_time_average}}

We need to show 
$(\overline{y_i}, \overline{y_t}) =(y_{i,Q},y_{t,Q})$ for $w=1$,
where
 $\overline{y_i} =\frac{1}{T}\int_0^T y_i dt$, $\overline{y_t} =\frac{1}{T}\int_0^T y_t dt$,
$T$ denotes the period of a cycle and 
$\mathbf{Q} = (y_{i,Q},y_{t,Q})$ is given by Eq.\,\eqref{eq:Q-coordinate}. By dividing both sides of Eq.\,\eqref{eq_replicator1} by $y_i(1-y_i) >0$ and substituting $w=1$, we obtain
$\frac{\dot{y}_i}{y_i(1-y_i)} =\left(\frac{r}{N_I}+1\right)y_t   -1 +v_I$.
Averaging both sides of the equation over time yields
$0 =\left(\frac{r}{N_I}+1\right)\overline{y_t} -1 +v_I$
since $\frac{1}{T}\int_0^T \frac{\dot{y}_i}{y_i(1-y_i)} dt =0$, which follows from $y_i(0)=y_i(T)$.
On the other hand, Eq.\,\eqref{eq:Q-coordinate} yields $\left(\frac{r}{N_I}+1\right)y_{t,Q} -1 +v_I = 0$.
Therefore, we obtain $\overline{y_t} =y_{t,Q}$.
Similarly, we can show $\overline{y_i} =y_{i,Q}$
by starting with dividing both sides of Eq.\,\eqref{eq_replicator2} by $y_t(1-y_t) >0$.

We need to show 
$\frac{1}{T}\int_0^T P  dt =P(\mathbf{Q})$,
where
$P (y_i, y_t)
=\frac{2 r N_I y_i y_t +N_I v_I y_i +N_T v_T y_t -a (N_I v_I+N_T v_T)}{N_I+N_T}
$.
Because we have shown $\overline{y_i} =y_{i,Q}$ and $\overline{y_t} =y_{t,Q}$ above,
we only need to show $\overline{y_i y_t} =\overline{y_i} \, \overline{y_t}$. 
To show this, we note that
$\frac{1}{y_iy_t}\frac{d (y_i y_t)}{dt} 
=y_i y_t   \left[\frac{N_I (1-r)}{N_T}-r-1\right] +y_i \left[\frac{N_I (r-1)}{N_T}-v_I+1\right] + y_t(r-v_T+1)+v_I+v_T-1$.
Averaging both sides of the equation over time yields 
$0 =\overline{y_i y_t}\left(\frac{N_I (1-r)}{N_T}-r-1\right) +\overline{y_i} \left(\frac{N_I (r-1)}{N_T}-v_I+1\right)  +\overline{y_t}(r-v_T+1) +v_I+v_T-1$
since $\frac{1}{T}\int_0^T \frac{1}{y_iy_t}\frac{d (y_i y_t)}{dt}  dt = 0$.
Therefore, we use $\overline{y_i} =y_{i,Q} =\frac{N_T v_T}{N_I(1-r)}$ and
$ \overline{y_t} =y_{t,Q}  =\frac{1-v_I}{1+r}$ to obtain
$\overline{y_i y_t} 
= \frac{\overline{y_i} (N_I (r-1)+N_T (1-v_I)) +\overline{y_t}N_T 
   (r-v_T+1)+N_T (v_I+v_T-1)}{r
   (N_I+N_T)-N_I+N_T}
=\frac{N_T v_T (1-v_I)}{N_I (1-r)(1+r)}
=\overline{y_i}\, \overline{y_t}$.

\subsection{Heteroclinic Cycle for $w>1$\label{heteroclinic_cycle}} 

Assume that $w>1$, $0 < v_T<v^*_T$ and $0\le v_I <1$. We first show that the heteroclinic cycle $\mathbf{F}_0 \equiv(0,0)\rightarrow \mathbf{F}_1 \equiv(0,1)\rightarrow \mathbf{F}_2 \equiv(1,1)\rightarrow \mathbf{F}_3 \equiv(1,0)\rightarrow \mathbf{F}_0$ is attracting, i.e., trajectories converge to it.
We obtain $\lambda_1|_\mathbf{y} >0$ and $\lambda_2|_\mathbf{y} <0$,
where $\lambda_1|_\mathbf{y}$ and $\lambda_2|_\mathbf{y}$ are eigenvalues of the Jacobian at $\mathbf{y} \in \{\mathbf{F}_0,\mathbf{F}_1,\mathbf{F}_2,\mathbf{F}_3 \}$.
Specifically, we obtain $\lambda_1|_{\mathbf{F}_0}= v_T$, $\lambda_1|_{\mathbf{F}_1} =r + v_I$,
$\lambda_1|_{\mathbf{F}_2}= \frac{(1 - r)(w^{N_I}-1) -N_T v_T(w-1)}{N_T (w-1)}$, 
$\lambda_1|_{\mathbf{F}_3}=1 - v_I$,
$\lambda_2|_{\mathbf{F}_0}= -1 + v_I$,
$\lambda_2|_{\mathbf{F}_1}=-v_T$, 
$\lambda_2|_{\mathbf{F}_2}= -\frac{r(w^{N_I}-1)  + N_I v_I (w-1) }{N_I (w-1)}$ and
$\lambda_2|_{\mathbf{F}_3}= -\frac{(1-r)(w^{N_I}-1) -N_T v_T(w-1)}{N_T (w-1)}$.
In other words,
each $\mathbf{y}$ is a saddle point.
The heteroclinic cycle $\mathbf{F}_0\rightarrow\mathbf{F}_1\rightarrow\mathbf{F}_2\rightarrow\mathbf{F}_3\rightarrow\mathbf{F}_0$ is attracting 
since $\rho \equiv \left(\frac{-\lambda_2|_{\mathbf{F}_0}}{\lambda_1|_{\mathbf{F}_0}}\right)\left(\frac{-\lambda_2|_{\mathbf{F}_1}}{\lambda_1|_{\mathbf{F}_1}}\right)\left(\frac{-\lambda_2|_{\mathbf{F}_2}}{\lambda_1|_{\mathbf{F}_2}}\right)\left(\frac{-\lambda_2|_{\mathbf{F}_3}}{\lambda_1|_{\mathbf{F}_3}}\right)
=\frac{r \left(w^{N_I}-1\right)+N_I v_I (w-1)}{N_I (w-1) (r+v_I)}  >1$,
according to the proof of Lemma 1 of Ref.\,\cite{Gaunersdorfer:1992wt}. 

We show $\frac{r \left(w^{N_I}-1\right)+N_I v_I (w-1)}{N_I (w-1) (r+v_I)}  >1$ as follows.
Using $w>1$, we obtain $1 + N_I (-1 + w) - w^{N_I} < 0$ since $\frac{\partial}{\partial w} \left[1 + N_I (-1 + w) - w^{N_I}\right] =N_I \left(1- w^{N_I -1}\right) <0$ and $\left[1 + N_I (-1 + w) - w^{N_I}\right] |_{w=1} =0$.
We then obtain $1 + N_I (-1 + w) - w^{N_I} < 0
\Longleftrightarrow r (1 + N_I (-1 + w) - w^{N_I}) < 0
\Longleftrightarrow N_I r (w-1) < r (w^{N_I}-1)
\Longleftrightarrow N_I r (w-1) +N_I v_I (w-1) < r (w^{N_I}-1) +N_I v_I (w-1)
\Longleftrightarrow \frac{r \left(w^{N_I}-1\right)+N_I v_I (w-1)}{N_I (w-1) (r+v_I)} >1$.

The time average $\frac{1}{T}\int_0^T (y_i,y_t) dt$ does not converge,
where $(y_i,y_t) =(y_i(t),y_t(t))$ is a trajectory converging to the heteroclinic cycle.
According to Theorem 1 of Ref.\,\cite{Gaunersdorfer:1992wt}, instead,
$\frac{1}{T}\int_0^T (y_i,y_t) dt$ asymptotically spirals towards the boundary of a polygon (i.e.\,a quadrangle) $\mathbf{A}_0\mathbf{A}_1\mathbf{A}_2\mathbf{A}_3$,
where $\mathbf{A}_i \equiv \frac{\mathbf{F}_{i+1} +\rho_{i+2}\mathbf{F}_{i+2} +\rho_{i+2}\rho_{i+3}\mathbf{F}_{i+3} +\rho_{i+2}\rho_{i+3}\rho_{i+4}\mathbf{F}_{i+4}}{1 +\rho_{i+2} +\rho_{i+2}\rho_{i+3} +\rho_{i+2}\rho_{i+3}\rho_{i+4}}$, 
$\rho_i \equiv \frac{-\lambda_2|_{\mathbf{F}_{i-1}}}{\lambda_1|_{\mathbf{F}_i}}$
and the indices are counted by modulo 4 (e.g.\,$\mathbf{F}_4 =\mathbf{F}_0$, $\rho_5 =\rho_1$).
Because the points $\mathbf{A}_i, \mathbf{A}_{i+1}$ (with $i \in \{1, 2, 3, 4 \}$) and $\mathbf{F}_{i+1}$ are collinear,
$\frac{1}{T}\int_0^T (y_i,y_t) dt$ asymptotically moves on a line from $\mathbf{A}_i$ to $\mathbf{A}_{i+1}$ in the direction of $\mathbf{F}_{i+1}$ in each cycle \cite{Gaunersdorfer:1992wt}.

Although the time averages of $y_i$ and $y_t$ do not converge,
the time average of the payoff $\overline{P} =\frac{1}{T}\int_0^T P dt $ converges. According to Lemma 1 of Ref.\,\cite{Gaunersdorfer:1992wt}, 
the time for which the trajectory spends near the saddle points $\mathbf{F}_i$ asymptotically grows $\rho$ times larger every cycle,
whereas the time required to move from a neighbourhood of one saddle point to that of the next one changes little. Thus, we can neglect the latter in comparison with the former.
Then, we obtain 
\begin{align}
\overline{P} =& \frac{t_0 P(\mathbf{F}_0) +t_1P(\mathbf{F}_1) +t_2P(\mathbf{F}_2) +t_3P(\mathbf{F}_3)}{t_0 +t_1 +t_2 +t_3}\notag \\
=& \frac{P(\mathbf{F}_0) +\frac{t_1}{t_0} P(\mathbf{F}_1) +\frac{t_2}{t_0}P(\mathbf{F}_2) +\frac{t_3}{t_0}P(\mathbf{F}_3)}{1 +\frac{t_1}{t_0}  +\frac{t_2}{t_0} +\frac{t_3}{t_0}} \notag\\
=& \frac{P(\mathbf{F}_0) +\frac{t_1}{t_0} P(\mathbf{F}_1) +\frac{t_1}{t_0}\frac{t_2}{t_1}P(\mathbf{F}_2)  +\frac{t_1}{t_0}\frac{t_2}{t_1}\frac{t_3}{t_2}P(\mathbf{F}_3)}{1 +\frac{t_1}{t_0}  +\frac{t_1}{t_0}\frac{t_2}{t_1} +\frac{t_1}{t_0}\frac{t_2}{t_1}\frac{t_3}{t_2}} \notag\\
=& \frac{P(\mathbf{F}_0) +\rho_1 P(\mathbf{F}_1) +\rho_1\rho_2 P(\mathbf{F}_2) +\rho_1\rho_2\rho_3 P(\mathbf{F}_3)}{1 +\rho_1  +\rho_1\rho_2  +\rho_1\rho_2\rho_3},
\label{eq:overlineP-heteroclinic}
\end{align}
where $t_i$ denotes the time for which the trajectory spends in an arbitrarily small neighbourhood of $\mathbf{F}_i$
and we have used $\frac{t_{i+1}}{t_i} =\rho_{i+1}$ from Lemma 1 of Ref.\,\cite{Gaunersdorfer:1992wt}.
Note that $\overline{P}_{\text{hc}}$ is a convex combination of $P(\mathbf{F}_0), P(\mathbf{F}_1), P(\mathbf{F}_2)$, and $P(\mathbf{F}_3)$.
By substituting Eq.~\eqref{eq:P(y_i,y_t)} with $(y_i, y_t) = (0, 0)$, $(0, 1)$, $(1, 0)$ and $(1, 1)$ in Eq.~\eqref{eq:overlineP-heteroclinic}, we obtain 
\begin{equation}
\begin{aligned}
\overline{P}_{\text{hc}} &=\left[\frac{1}{(w-1) \left\{\frac{(r+1) \left[ N_I (1-r) + r N_Tv_T\right]}{N_T v_T \left[r
   \left(w^{N_I}-1\right)+N_I (w-1)\right]}-\frac{r}{w^{N_I}-1}\right\}} \right. \\
   & \quad \left. -a (N_I v_I+N_T v_T)\right] \frac{1}{N_I+N_T}. 
\end{aligned}
\end{equation}

\subsection{Optimal Incentives\label{optimal_incentive}} 

In this section, we calculate the optimal incentive and payoff when
$w\le 1$ and when $\left(w>1 \land N_I \rightarrow \infty\right)  \lor (w \rightarrow \infty)$.

\subsubsection{$w\le 1$} 

To obtain the optimal payoff, we need to know $\max \{P^*(0,0), P^*(1,0), P^*(1,1), P^*(\mathbf{Q})\}$.
We obtain $P^*(1,1) -P^*(\mathbf{Q})=\frac{r\left(w^{N_I}-1\right)}{(N_I+N_T)(w-1)} +\frac{(a-1) N_T}{N_IN_T}\epsilon >0$.
In addition, we have 
\begin{align}
\triangle P_1 \equiv& P^*(1,1) -P^*(1,0) \notag\\
=& \frac{[(a+1) r-a] \left(w^{N_I}-1\right)}{(N_I+N_T)(w-1)}+\frac{a
   N_I}{N_I+N_T}+(a-1) \epsilon \notag\\
>& 0
\end{align}
for $0< w \le 1$ because $\triangle P_1$ is a monotonic function of $w>0$, we have $\triangle P_1 |_{w=0} = \frac{a (N_I+r-1)+r}{N_I+N_T} +(a-1)\epsilon  >0$, and we have $\triangle P_1|_{w=1} =\frac{(a+1)rN_I}{N_I+N_T}  +(a-1)\epsilon   >0$.
Therefore, it holds true that $\max \{P^*(0,0), P^*(1,0), P^*(1,1), P^*(\mathbf{Q})\} =\max \{P^*(0,0), P^*(1,1)\}$.

If $r > r^*_1 = \frac{a-1}{a+1}$,
then 
\begin{equation} 
\begin{aligned}
\triangle P_2 & \equiv P^*(1,1) -P^*(0,0) \\
 & =\frac{(a+1) r-a+1}{N_I+N_T}\frac{w^{N_I}-1}{w-1} +(1-a)\epsilon >0.
\end{aligned}
\label{eq:P^*(1,1)-P^*(0,0)>0}
\end{equation}
In this case, the optimal payoff is $P^*(1,1)$, and the corresponding optimal incentive is $v_I=0\land v_T=v^*_T +\epsilon$.
If $r < r^*_1$, then $\triangle P_2 <0$. In this case,
the optimal payoff is $P^*(0,0)$, and the corresponding optimal incentive is $v_I=0\land v_T=0$.

\subsubsection{$\left(w>1 \land N_I \rightarrow \infty\right) \lor (w \rightarrow \infty)$\label{optimal_incentive_w>1}} 

As $N_I \rightarrow \infty$, we obtain
\begin{equation}
\triangle P_1 = P^*(1,1) -P^*(1,0)
\rightarrow \frac{(a+1) r-a}{N_I+N_T} \frac{w^{N_I}-1}{w-1}.
\label{eq:P^*(1,1)>P^*(1,0)}
\end{equation}
The sign of $\triangle P_1$ is determined by that of $(a+1) r-a$ since $\frac{w^{N_I}-1}{w-1} >0$.
Therefore, if $r >r^*_2=\frac{a}{a+1}$, then $P^*(1,1) >P^*(1,0)$, and if $r <r^*_2$, then $P^*(1,1) <P^*(1,0)$.
As $N_I \rightarrow \infty$, we also obtain
\begin{equation}
P^*(1,0) -\overline{P}^*_{\text{hc}} \rightarrow \infty,
\label{eq:a-diff-diverges}
\end{equation}
where we remind that $\overline{P}^*_{\text{hc}}$ denotes the maximum of $\overline{P}_{\text{hc}}$ with respect to $v_I$ and $v_T$.
We prove Eq.~\eqref{eq:a-diff-diverges} in Appendix~\ref{app:proof-54}.

Equations~\eqref{eq:P^*(1,1)>P^*(1,0)} and \eqref{eq:a-diff-diverges} imply the following.
First, if $r>r^*_2$, then $\max \{P^*(0,0), P^*(1,1), P^*(1,0), \overline{P}^*_{\text{hc}}\} =\max \{P^*(0,0), P^*(1,1)\}$.
Since $r>r^*_1$, which follows from
$r^*_2 >r^*_1$, we obtain $\triangle P_2 =P^*(1,1) -P^*(0,0) >0$, which we showed in
Eq.~\eqref{eq:P^*(1,1)-P^*(0,0)>0}.
Therefore, $\max \{P^*(0,0), P^*(1,1)\} =P^*(1,1)$;
$P^*(1,1)$ is the optimal payoff, and the associated optimal incentive is $v_I=0 \land v_T=v^*_T +\epsilon $.
Second, if $r <r^*_2$, then $\max \{P^*(0,0), P^*(1,1), P^*(1,0), \overline{P}^*_{\text{hc}}\} =\max \{P^*(0,0), P^*(1,0)\}$. In this case, we obtain
$\triangle P_3 \equiv P^*(1,0) -P^*(0,0) =\frac{1}{N_I+N_T} \left(-N_Ia +\frac{1-w^{N_I}}{1-w}\right) -\frac{(a-1) N_I}{N_I+N_T} \epsilon \rightarrow \infty$. Therefore, $P^*(1,0)$ is the optimal payoff, and the associated optimal incentive is $v_I=1 +\epsilon \land v_T=0$.

Finally, when $N_I$ is finite and $w \rightarrow \infty$, we have the same outcome via similar calculations.

\subsection{Proof of Eq.~\eqref{eq:a-diff-diverges}\label{app:proof-54}}

To prove Eq.~\eqref{eq:a-diff-diverges}, we first show that $\overline{P}_{\text{hc}}$ is monotonic or has a local maximum as a function of $v_T \in (0, v_T^*)$.
Since the denominator of
$\frac{\partial \overline{P}_{\text{hc}}}{\partial v_T}$ is 
$(w-1) \left\{N_I \left[(1-r^2) \left(w^{N_I}-1\right) -N_T r v_T (w-1)\right] \right. \\ \left. +N_T r v_T \left(w^{N_I}-1\right)\right\}^2
   >0$,
the sign of $\frac{\partial \overline{P}_{\text{hc}}}{\partial v_T} $ is determined by that of its numerator,
$c(v_T) \equiv 
   -v_T^2 N_T^3 a r^2  (w-1) \left(w^{N_I}-N_I w+N_I-1\right)^2
\\+2v_T N_I N_T^2 a r (w-1)\left(w^{N_I}-1\right)\left(r^2-1\right)\left[\left(w^{N_I}-1\right) \right. \\ \left.- N_I (w-1) \right]
   \\
   +N_T N_I \left(1-r^2\right)\left(w^{N_I}-1\right)^2 \left[N_I  (w-1)\left(a r^2-a+1\right) \right. \\ \left. + r \left(w^{N_I}-1\right)\right]
$,
which is a quadratic equation of $v_T$.
Of the two real solutions of $c(v_T)=0$, we consider only the larger one,
$v_T =v^{\text{hc}}_T = \frac{ \left\{\sqrt{aN_I \left(1-r^2\right) (w-1) \left[r \left(w^{N_I}-1\right)+N_I
   (w-1)\right]} -aN_I \left(1-r^2\right) (w-1)\right\}}{a N_T r (w-1) \left(w^{N_I}-N_I w+N_I-1\right)} \times \left(w^{N_I}-1\right)$
because the smaller one is guaranteed to be always negative and $v_T \ge 0$.
Since the coefficient of the quadratic term $v^2_T$ is negative,
the sign of $c(v_T)$ over the domain $(0,v^*_T)$ can be entirely positive, entirely negative, or change from positive to negative just once.
In other words, $\overline{P}_{\text{hc}}$  is monotonic or has a local maximum as a function of $v_T \in (0,v^*_T)$.
The local maximum of $\overline{P}_{\text{hc}}$, if it exists, is realized at $v_T =v^{\text{hc}}_T$.
Because $\frac{\partial \overline{P}_{\text{hc}}}{\partial v_I} =-\frac{a N_I}{N_I+N_T} <0$ implies that
the maximum of $\overline{P}_{\text{hc}}$ in terms of $v_I$ is realized at $v_I =0$ regardless of the value of $v_T$,
we conclude that $\overline{P}_{\text{hc}}^*$ is equal to either $\overline{P}_{\text{hc}}(0 +\epsilon)$, $\overline{P}_{\text{hc}}(v^{\text{hc}}_T)$ or $\overline{P}_{\text{hc}}(v^*_T -\epsilon)$.

We obtain 
$P^*(1,0) -\overline{P}_{\text{hc}}(0 + \epsilon) 
= \frac{1-w^{N_I}}{(1-w)   (N_I+N_T)} 
-\frac{a   N_I}{N_I+N_T}
+\frac{\left[(1-a) N_I+a N_T\right]}{N_I+N_T}\epsilon  
   -\frac{1}{(w-1) (N_I+N_T) \left\{ \frac{(r+1)
   \left[N_I (1-r) +N_T r \epsilon \right]}{N_T \epsilon  \left[r
   \left(w^{N_I}-1\right)+N_I
   (w-1)\right]}-\frac{r}{w^{N_I}-1}\right\}}
    \rightarrow\infty$ as $N_I \rightarrow \infty$ 
    and $\epsilon \rightarrow 0$.
At $v_T = v^{\text{hc}}_T$, we have 
$P^*(1,0) -\overline{P}_{\text{hc}}(v^{\text{hc}}_T)  
= \frac{a N_I^2 r (w-1)^2 +2 \left(w^{N_I}-1\right) \sqrt{a N_I \left(1-r^2\right) (w-1) \left[r
   \left(w^{N_I}-1\right)+N_I (w-1)\right]}}{r (w-1)
   (N_I+N_T) \left(w^{N_I}-N_I w+N_I-1\right)} 
+\frac{N_I \{a [(r-1) r-1]-r-1\} \left(w^{N_I}-1\right)}{r 
   (N_I+N_T) \left(w^{N_I}-N_I w+N_I-1\right)} +\frac{a N_I}{N_I+N_T} \epsilon 
   \rightarrow \infty$  
as $N_I \rightarrow \infty$ and $\epsilon \rightarrow 0$.
Finally, we have 
$P^*(1,0) -\overline{P}_{\text{hc}}(v^*_T - \epsilon)
= \frac{\left[a (1-r)+1\right] \left(w^{N_I}-1\right)}{(w-1) (N_I+N_T)} -\frac{aN_I}{N_I +N_T} +\frac{(1-a) N_I-a N_T}{N_I+N_T} \epsilon    
   -\frac{1}{\frac{(r+1)
   \left\{ r \left[(r-1) w^{N_I}+N_T w \epsilon -N_T \epsilon
   -r+1\right]+N_I (r-1) (w-1)\right\}}{\left[r \left(w^{N_I}-1\right)+N_I
   (w-1)\right] \left[(r-1) w^{N_I}+N_T w \epsilon -N_T \epsilon
   -r+1\right]}-\frac{r}{w^{N_I}-1}}  \times \frac{1}{(w-1) (N_I+N_T) }$, which tends to
$\frac{2 \left(w^{N_I}-1\right) \sqrt{r \left(w^{N_I}-1\right)+N_I (w-1)}   }{r (w-1) (N_I+N_T)   \left(w^{N_I}-N_I w+N_I-1\right)} \times \sqrt{-a N_I \left(r^2-1\right) (w-1)}
   -\frac{a N_I \left[N_I   (1-w) +r-1\right]}{(N_I+N_T) \left(w^{N_I}-N_I   w+N_I-1\right)} 
   +\frac{a N_I \left([(r-1) r-1] w^{N_I}+1\right)}{r   (N_I+N_T) \left(w^{N_I}-N_I w+N_I-1\right)}
   -\frac{N_I   (r+1) \left(w^{N_I}-1\right)}{r (N_I+N_T) \left(w^{N_I}-N_I   w+N_I-1\right)}$ as $\epsilon \rightarrow 0$. As $N_I \rightarrow \infty$, this quantity tends to
$\frac{2 w^{\frac{N_I-1}{2}} \sqrt{a N_I \left(1-r^2\right)}}{\sqrt{r} (N_I+N_T)} \rightarrow \infty$. 
This concludes the proof of Eq.~\eqref{eq:a-diff-diverges}.

\bibliographystyle{IEEEtran}    

%\bibliography{/Users/iksoolim/Google_Drive/Bibliography-link/graphics}  
%\bibliography{/Users/iksoolim/GoogleDriveNoSpace/Bibliography-link/graphics}  


\end{document}